\documentclass[12pt, letterpaper]{article}

\usepackage{amssymb}
\usepackage{amsmath}
\usepackage{verbatim}

\usepackage[nosort]{cite}
\usepackage[hyperref,bulletsep]{collect}

\mathversion{normal}

\makeatletter \@addtoreset{equation}{section} \makeatother

\newcommand{\gen}[1]{\mathfrak{#1}}
\newcommand{\alg}[1]{\mathfrak{#1}}

\ifx\genfrac\sdflkaj

\else

\fi
\newcommand{\sfrac}[2]{{\textstyle\frac{#1}{#2}}}
\ifx\half\sdflkaj
\newcommand{\half}{\sfrac{1}{2}}
\fi
\newcommand{\quarter}{\sfrac{1}{4}}

\newcommand{\comm}[2]{[#1,#2]}
\newcommand{\acomm}[2]{\{#1,#2\}}

\newcommand{\gcomm}[2]{[#1,#2\}}

\newcommand{\state}[1]{\mathopen{\big|}#1\mathclose{\bigr \rangle}}

\let\oldPhi=\Phi
\let\oldPsi=\Psi
\renewcommand{\Phi}{\mathnormal{\oldPhi}}
\renewcommand{\Psi}{\mathnormal{\oldPsi}}

\newcommand{\nln}{\nonumber\\}
\newcommand{\nl}{\nonumber\\&&\mathord{}}
\newcommand{\earel}[1]{\mathrel{}&#1&\mathrel{}}
\newcommand{\eq}{\earel{=}}
\newenvironment{myeqnarray}{\arraycolsep0pt\begin{eqnarray}}{\end{eqnarray}\ignorespacesafterend}
\newenvironment{myeqnarray*}{\arraycolsep0pt\begin{eqnarray*}}{\end{eqnarray*}\ignorespacesafterend}

\def\[{\begin{equation}}
\def\]{\end{equation}}
\def\<{\begin{myeqnarray}}
\def\>{\end{myeqnarray}}

\ifx\href\asklfhas\newcommand{\href}[2]{#2}\fi
\newcommand{\arxivno}[1]{\href{http://arxiv.org/abs/#1}{#1}}

\begin{document}
\thispagestyle{empty}
\begin{flushright}\footnotesize
\texttt{\arxivno{arXiv:0901.0411}}\\ 
\texttt{DAMTP-2009-1}
\end{flushright}
\vspace{.5cm}

\begin{center}%
{\Large\textbf{\mathversion{bold}%
Two-loop Integrability of Planar \\ $\mathcal{N}=6$ Superconformal  Chern-Simons Theory }\par}\vspace{1.5cm}%

\textsc{Benjamin~I.~Zwiebel} \vspace{8mm}

\textit{ DAMTP, Centre for Mathematical Sciences \\ 
University of Cambridge\\%
 Wilberforce Road, Cambridge CB3 0WA, UK}%
\vspace{4mm}

\texttt{b.zwiebel@damtp.cam.ac.uk}\par\vspace{1.5cm}

\textbf{Abstract}\vspace{7mm}

\begin{minipage}{14.7cm}
Bethe ansatz equations have been proposed for the asymptotic spectral problem of $AdS_4/CFT_3$. This proposal assumes integrability, but the previous verification of weak-coupling integrability covered only the $\alg{su}(4)$ sector of the ABJM gauge theory. Here we derive the complete planar two-loop   dilatation generator of $\mathcal{N}=6$ superconformal  Chern-Simons theory from $\alg{osp}(6|4)$ superconformal symmetry. For the $\alg{osp}(4|2)$ sector, we prove integrability through a Yangian construction. We argue that integrability extends to the full planar two-loop dilatation generator, confirming the applicability of the Bethe equations at weak coupling. Further confirmation follows from an analytic computation of the two-loop twist-one spectrum.
\end{minipage}

\end{center}

\newpage
\setcounter{page}{1}
\renewcommand{\thefootnote}{\arabic{footnote}}
\setcounter{footnote}{0}



\section{Introduction}
The $\mathcal{N}=6$ superconformal $U(N) \times U(N)$ Chern-Simons theory of Aharony, Bergman, Jafferis, and Maldacena has a 't Hooft limit in which the dual description reduces to type IIA string theory on $AdS_4 \times CP^3$ \cite{Aharony:2008ug}.  This limit is $N \to \infty$ and  Chern-Simons level $k \to \infty$ with coupling $\lambda=N/k$ held fixed. Excitingly, the asymptotic spectral problem of this $AdS_4/CFT_3$ duality may be solvable by an $\alg{osp}(6|4)$ Bethe ansatz, as first argued by Minahan and Zarembo based on their two-loop ($\mathcal{O}(\lambda^2)$) gauge theory calculation\cite{Minahan:2008hf} (also see \cite{Bak:2008cp}). Paralleling the case of $AdS_5/CFT_4$ \cite{Minahan:2002ve,Beisert:2003tq,Bena:2003wd,Arutyunov:2004vx,Staudacher:2004tk, Beisert:2005tm, Janik:2006dc,Arutyunov:2006ak,Beisert:2006ez}, this apparent integrability is likely to lead to a better understanding of $AdS_4/CFT_3$ and the relationship between gauge/string duality and integrability.

There has been much related activity. Integrability of classical string theory on $AdS_4 \times CP^3$ has been proved for a subsector described by an $OSP(6|4)/(U(3) \times SO(1,3))$ coset sigma model \cite{Arutyunov:2008if,Stefanski:2008ik}.  The conjectured weak-coupling Bethe equations \cite{Minahan:2008hf} have been extended to an all-loop proposal \cite{Gromov:2008qe}\footnote{The all-loop proposal depends on an unknown interpolating function that appears in the one-magnon dispersion relation \cite{Nishioka:2008gz,Gaiotto:2008cg},\cite{Grignani:2008is,Berenstein:2008dc}.}, with the corresponding two-magnon S-matrix obtained in \cite{Ahn:2008aa}.  This all-loop Bethe ansatz has passed strong-coupling tests in a $SU(2) \times SU(2)$ sector\cite{Astolfi:2008ji,Sundin:2008vt}, and seeming disagreement with  semi-classical string computations \cite{McLoughlin:2008ms,Alday:2008ut,Krishnan:2008zs} is eliminated by a one-loop correction \cite{McLoughlin:2008he} to the magnon dispersion relation. Nonetheless, discrepancies between one-loop strong-coupling string calculations and the Bethe ansatz remain \cite{McLoughlin:2008he}, and the classical integrability of the \emph{complete} $AdS_4 \times CP^3$ Green-Schwarz action \cite{Gomis:2008jt} is unverified.

This work, however, will focus on the weak-coupling planar gauge theory. As is the case for $\mathcal{N}=4$ SYM, the spectral problem of computing anomalous dimensions is usefully formulated in terms of a spin chain. Local gauge-invariant operators are mapped to spin-chain states, and the anomalous part of the dilatation generator is mapped to a spin-chain Hamiltonian whose eigenvalues are the spectrum of anomalous dimensions. Then, if this Hamiltonian is integrable, one can replace the eigenvalue problem with a dramatically simpler problem of solving a system of Bethe equations, see \cite{Beisert:2004ry} for a review. 

For the ABJM gauge theory, \cite{Minahan:2008hf} calculated the leading-order two-loop planar dilatation generator or Hamiltonian in the $\alg{su}(4)$ sector, confirming two-loop integrability within that sector.  The $\alg{su}(4)$ sector corresponds to local operators composed of alternating scalars $\phi_i$ and $\bar{\phi}^i$, $i=1,\ldots4$, or a spin chain with alternating sites occupied by a spin transforming in the $\mathbf{4}$ or $\mathbf{\bar{4}}$ of $\alg{su}(4)$. The spin chain is alternating because the $\phi_i$ transform as $(\mathbf{N}, \mathbf{\bar{N}})$  with respect to the $U(N) \times U(N)$ gauge group, while the $\bar{\phi}^i$ transform as $(\mathbf{\bar{N}},\mathbf{N})$.  

To describe the full set of local operators, we must extend these conjugate finite-dimensional representations of the $\alg{su}(4) = \alg{so}(6)$ $R$ symmetry to two highest-weight irreducible representations of the $\mathcal{N}=6$ superconformal Lie algebra, $\alg{osp}(6|4)$. In addition to the $\alg{su}(4)$ scalars and their supersymmetric partners, $\bar{\psi}^i$ or  $\psi_i$, these representations or modules includes arbitrarily many (symmetrized traceless) covariant derivatives acting on each scalar or fermion. The validity of the weak-coupling $\alg{osp}(6|4)$ Bethe equation conjecture requires the significant assumption that two-loop integrability extends to the alternating spin chain with sites hosting these infinite-dimensional $\alg{osp}(6|4)$ modules.  

Before working on extending the two-loop $\alg{su}(4)$ sector dilatation generator to the full $\alg{osp}(6|4)$ spin chain, it is useful to review key steps used to compute the leading-order one-loop planar dilatation generator of $\mathcal{N}=4$ SYM.  For that theory, the one-loop dilatation generator acts on two (identical) $\alg{psu}(2,2|4)$ modules. An important observation is that the irreducible multiplets in the tensor product of two modules match up one-to-one with the corresponding multiplets in the $\alg{sl}(2)$ sector\footnote{$\mathcal{N}=4$ SYM actually has two types of $\alg{sl}(2)$ sectors, with bosonic and fermionic module elements respectively. For our purposes it is not important which sector was used for each of the following steps, so we will not again refer to this distinction.}. Combined with superconformal invariance, this enabled the lift of the $\alg{sl}(2)$ sector dilatation generator\footnote{The one-loop $\alg{sl}(2)$ sector dilatation generator is fixed by the twist-two spectrum, which was also computed earlier in \cite{Lipatov:1997vu,Dolan:2001tt}.} to the complete one-loop dilatation generator \cite{Beisert:2003jj}. In fact,  superconformal symmetry completely fixes the leading-order Hamiltonian (up to normalization); the $\alg{sl}(2)$ sector one-loop dilatation generator was later computed \cite{Beisert:2004ry} just by requiring closure of the residual superconformal symmetry algebra of the sector, which includes an extra $\alg{su}(1|1)$ symmetry generated by  ``hidden'' supercharges.

We follow a similar path for the planar ABJM theory, using superconformal symmetry directly\footnote{As in the $\mathcal{N}=4$ SYM calculations, we also use basic structural properties of the gauge theory, such as the range of planar interactions.} to compute the leading-order dilatation generator up to normalization\footnote{An alternative approach would be to use the construction of \cite{Sochichiu:2008tw} of the two-loop dilatation generator for arbitrary three-dimensional Lorentz-invariant renormalizable theories. That is likely to be a much more difficult approach since it would not take full advantage of superconformal symmetry.}. Again the first step involves restricting to a sector; we find the leading-order $\alg{osp}(4|2)$ sector dilatation generator using residual superconformal symmetry.  This sector corresponds to the set of local operators that are $1/12$ BPS at $\lambda=0$ with respect to a fixed pair of supercharges, which we label $\hat{\gen{Q}}$ and $\hat{\gen{S}}$. Similar to the case of the $\alg{sl}(2)$ sector of $\mathcal{N}=4$ SYM, these supercharges generate an important additional $\alg{su}(1|1)$, which commutes with $\alg{osp}(4|2)$. By definition, all states in this sector are annihilated by these supercharges at $\lambda=0$, so we then have
\[
\delta \gen{D}_2 =  \acomm{\hat{\gen{Q}}_1}{\hat{\gen{S}}_1}, \label{eq:introd2}
\]
where subscripts refer to the order in $\lambda$. In other words, the two-loop dilatation generator within this sector is fixed once we know the \emph{one-loop} corrections to these supercharges. Furthermore, requiring vanishing commutators with  $\alg{osp}(4|2)$ generators at leading order fixes $\hat{\gen{Q}}_1$ and $\hat{\gen{S}}_1$ up to overall normalization, giving the two-loop dilatation generator (the normalization can be identified from \cite{Minahan:2008hf}). 

Superconformal invariance enables us to lift the leading-order dilatation generator within the $\alg{osp}(4|2)$ sector to the full theory. However, in comparison to the analogous $\mathcal{N}=4$ SYM calculation, additional intermediate steps are needed. This is largely because here the leading-order Hamiltonian acts on three adjacent sites. The expression that immediately follows from the right side of (\ref{eq:introd2}) is not in a convenient form for the lift. Instead we first prove integrability within this sector by constructing an $\alg{osp}(4|2)$ Yangian that commutes with the Hamiltonian (The construction of the Yangian symmetry of the one-loop dilatation generator of $\mathcal{N}=4$ SYM was completed in \cite{Dolan:2003uh,Dolan:2004ps}).  This identifies the Hamiltonian as one that follows from an $\alg{osp}(4|2)$ R-matrix construction. Now the Hamiltonian is written in terms of projectors onto irreducible modules appearing in the tensor products of two one-site modules, and now the lift becomes straightforward. Superconformal invariance, combined with some analysis of tensor products of one-site modules, allows us to lift this second expression for the $\alg{osp}(4|2)$ Hamiltonian to the full two-loop $\alg{osp}(6|4)$ spin-chain Hamiltonian. 

In this work we do not explicitly verify that the complete Hamiltonian has an  $\alg{osp}(6|4)$ Yangian symmetry. However, it takes the exact form of an integrable alternating $\alg{osp}(6|4)$ spin-chain Hamiltonian, assuming the existence of an  $\alg{osp}(6|4)$ R-matrix (This is parallel to a result of  \cite{Beisert:2003yb} for $\mathcal{N}=4$ SYM).  Combined with the explicit proof within the $\alg{osp}(4|2)$ sector, this is convincing evidence that the complete two-loop planar model is integrable and that the two-loop spectral problem is solved by the Bethe ansatz.

In Section \ref{sec:alg} we introduce the ABJM $\alg{osp}(6|4)$ spin-chain model and the restriction to the $\alg{osp}(4|2)$ sector, and Section \ref{sec:oneloop} derives the two-loop dilatation generator for this sector. Here we also compute analytically the twist-one spectrum from the dilatation generator, which matches the Bethe ansatz prediction. The Yangian proof of integrability appears in Section \ref{sec:yangian}, and the following section computes the corresponding R-matrix expression for the Hamiltonian. Section \ref{sec:lift} gives the unique lift to the complete two-loop planar dilatation generator, and we conclude with the following section.

\section{The $\alg{osp}(6|4)$ spin chain and its $\gen{osp}(4|2)$ subsector \label{sec:alg}}
The ABJM gauge theory's $\mathcal{N}=6$ superconformal symmetry, which has been verified in \cite{Benna:2008zy,Bandres:2008ry}, corresponds to the $\alg{osp}(6|4)$ Lie algebra. After reviewing this algebra we introduce the corresponding module that is used for building the spin-chain description of gauge-invariant local operators. We then explain the restriction to the $\alg{osp}(4|2)$ sector, and this sector's algebra and spin module. Finally, we introduce a light-cone superspace basis for the $\alg{osp}(4|2)$ module. For a recent general analysis of representations of the three dimensional superconformal groups, see \cite{Dolan:2008vc}.

\subsection{The complete algebra and spin module}
For the $\alg{osp}(6|4)$ algebra generators we use one $\alg{su}(2)$ spinor index, $\alpha, \beta = 1, 2$ and an $\alg{su}(4)$ index $i, j=1, 2, 3, 4$. The  $\alg{sp}(4)$ elements are $\alg{su}(2)$ Lorentz generators $\gen{L}^\alpha{}_\beta$, translation and special conformal generators  $\gen{P}_{\alpha\beta}=\gen{P}_{\beta\alpha}$ and  $\gen{K}^{\alpha\beta}=\gen{K}^{\beta\alpha}$, and the dilatation generator $\gen{D}$. The remaining bosonic elements are $\alg{su}(4)$    $\gen{R}$-symmetry generators $\gen{R}^i{}_j$. 
Additionally, there are 24 supercharges with
\[
\gen{Q}_{ij,\alpha} = - \gen{Q}_{ji,\alpha}, \quad \gen{S}^{kl,\beta} = - \gen{S}^{lk,\beta}.
\]
The nonvanishing commutators with the $\alg{sp}(4)$ generators are
\begin{align}
\comm{\gen{L}^\alpha{}_\beta}{\gen{L}^\gamma{}_\delta} &= \delta^\alpha_\delta \gen{L}^\gamma{}_\beta - \delta^\gamma_\beta \gen{L}^\alpha{}_\delta, & \comm{\gen{L}^\alpha{}_\beta}{\gen{P}_{\gamma\delta}}& =  2 \delta^\alpha_{\{\gamma} \gen{P}_{\delta\}\beta} - \delta^\alpha_\beta \gen{P}_{\gamma\delta},  
\notag \\
 \comm{\gen{L}^\alpha{}_\beta}{\gen{K}^{\gamma\delta}}& =  -2 \delta_\beta^{\{\gamma} \gen{K}^{\delta\}\alpha} + \delta^\alpha_\beta \gen{K}^{\gamma\delta}, &
\comm{\gen{L}^\alpha{}_\beta}{\gen{Q}_\gamma} &= \delta^\alpha_\gamma \gen{Q}_\beta -\half \delta^\alpha_\beta \gen{Q}_\gamma,  
\notag \\
\comm{\gen{L}^\alpha{}_\beta}{\gen{S}^\gamma} & =  -\delta^\gamma_\beta \gen{S}^\alpha +\half \delta^\alpha_\beta \gen{S}^\gamma, & \comm{\gen{K}^{\alpha \beta}}{\gen{P}_{\gamma\delta}} & =  4\delta^{\{\alpha}_{\{\gamma} \gen{L}^{\beta\}}{}_{\delta\}} + 4\delta^{\{\alpha}_{\{\gamma} \delta^{\beta\}}_{\delta\}} \gen{D}, 
\notag \\
\comm{\gen{K}^{\alpha \beta}}{\gen{Q}_{kl, \gamma}} & =  \half \varepsilon_{klij}\big(\delta^\alpha_\gamma \gen{S}^{ij, \beta} + \delta^\beta_\gamma \gen{S}^{ij, \alpha} \big), & &
\notag \\
\comm{\gen{P}_{\alpha \beta}}{\gen{S}^{kl,\gamma}} & =  -\half \varepsilon^{klij}\big(\delta_\alpha^\gamma \gen{Q}_{ij,\beta} + \delta_\beta^\gamma \gen{Q}_{ij,\alpha} \big), &&
\end{align}
and the dimensions of $\{\gen{P}, \gen{K}, \gen{Q}, \gen{S}\}$ are $\{1,-1,\half,-\half\}$.

The commutators with the  $\alg{so}(6) =  \alg{su}(4)$ $\gen{R}$  generators are
\begin{align}
\comm{\gen{R}^i{}_j}{\gen{R}^k{}_l} & = \delta^i_l\gen{R}^k{}_j -  \delta^k_j\gen{R}^i{}_l,& \comm{\gen{R}^i{}_j}{\gen{Q}_{kl}} &=2 \delta^i_{[l} \gen{Q}_{k]j}-\half \delta^i_j \gen{Q}_{kl}, \notag \\
\comm{\gen{R}^i{}_j}{\gen{S}^{kl}} & = -2 \delta_j^{[l} \gen{S}^{k]i}+\half \delta^i_j \gen{S}^{kl}. & &
\end{align}
Finally, the nonvanishing anticommutators are
\begin{align}
\acomm{\gen{Q}_{ij,\alpha}}{\gen{Q}_{kl,\beta}} & =  -\varepsilon_{ijkl} \gen{P}_{\alpha\beta},  &
\acomm{\gen{S}^{ij,\alpha}}{\gen{S}^{kl,\beta}} & =  -\varepsilon^{ijkl} \gen{K}^{\alpha\beta}, 
\notag \\
\acomm{\gen{Q}_{ij,\beta}}{\gen{S}^{kl,\gamma}} & =   4 \delta^\gamma_\beta \delta^{[k}_{[i}\gen{R}^{l]}{}_{j]}  + 2 \delta^k_{[j}\delta^l_{i]}\gen{L} ^\gamma{}_\beta + 2 \delta^k_{[j}\delta^l_{i]}\delta^\gamma_\beta \gen{D}.  && \label{eq:anticommutators}
\end{align}

There are multiple choices of positive roots of the Lie algebra. For instance, we can choose the raising generators as
\[
\gen{L}^2{}_1, \quad \gen{K}, \quad \gen{R}^i{}_j,  \, i > j,  \quad \gen{S}.
\]
and the Hermitian conjugate lowering generators
\[
\gen{L}^1{}_2,  \quad \gen{P}, \quad \gen{R}^i{}_j,  \, i < j, \quad \gen{Q}.
\]
The Cartan generators are then the diagonal generators of $\gen{L}$ and $\gen{R}$ (the traceless conditions mean that only $1+3$ of these are independent) and the dilatation generator.

The spin chain has alternating highest-weight modules. As stated in the introduction, this is because the matter fields are in bifundamental representations of the $U(N) \times U(N)$ gauge group. We also note that ABJM gauge theory has an additional $\gen{u}(1)$ symmetry, under which the modules have alternating charge $\pm 1$.   The first representation $\mathcal{V}_\phi$ has highest-weight element $\state{\phi_1^{(0,0)}}$ and consists of 
\[
\state{\phi_i^{(n_1, n_2)}}, \quad \state{(\bar{\psi}^i)^{(n_1, n_2)}}, \quad n_i  =0, 1, 2, \ldots
\]
where the superscripts correspond to the number of $\alg{su}(2)$ indices carried by covariant derivatives acting on the fields of the ABJM model, $\phi$ and $\bar{\psi}$. All Lorentz indices are symmetrized on individual module elements, as needed for an irreducible representation. Alternate sites of the chain, instead host a representation $\mathcal{V}_{\bar \phi}$ with highest-weight element $\state{(\bar{\phi}^4)^{(0,0)}}$ spanned by
\[
\state{(\bar{\phi}^i)^{(n_1, n_2)}}, \quad \state{\psi_i^{(n_1, n_2)}}, \quad n_i  =0, 1, 2, \ldots
\]
The $\alg{su}(4)$ sector corresponds to the $\phi$ and $\bar{\phi}$ states with all $n_i=0$. Also, of course, $n_1+n_2$ is even (odd) for bosons (fermions). 

The $\alg{osp}(6|4)$ variations were given in \cite{Gaiotto:2008cg,Hosomichi:2008jb}. We will not need the explicit action of the entire $\alg{osp}(6|4)$ generators at leading order, which can be identified (up to a physically irrelevant choice of basis) straightforwardly by requiring closure of the algebra. However, a few details are useful. The diagonal Lorentz generators and the classical dilatation generator act (independently of $\gen{R}$ symmetry indices) as
\<
\gen{L}^1{}_1\state{X^{(n_1, n_2)}} \eq -\gen{L}^2{}_2\state{X^{(n_1, n_2)}} = \half(n_1-n_2) \state{X^{(n_1, n_2)}}, 
\nln
\gen{D}_0 \state{X^{(n_1, n_2)}} \eq \half(n_1+n_2+1)\state{X^{(n_1, n_2)}}. \label{eq:lorentanddilatationaction}
\>
All unbarred elements transform in the $\mathbf{4}$ of $\alg{su}(4)$ and barred elements transform (naturally) in the $\mathbf{\bar{4}}$, 
\[ \label{eq:rsymaction}
\gen{R}^i{}_j \state{X_k} = \delta^i_k \state{X_j} - \quarter \delta^i_j \state{X_k}, \quad \gen{R}^i{}_j \state{\bar{X}^k} = -\delta^k_j \state{\bar{X}^i} + \quarter \delta^i_j \state{\bar{X}^k}.
\]
Also, schematically we have the $\gen{R}$-symmetry index dependence of the leading-order supercharges,
\begin{align}
\gen{Q}_{ij}\state{X_k} & \sim \varepsilon_{ijkm}\state{\bar{Y}^m}, &  \gen{Q}_{ij}\state{\bar{X}^k} & \sim \delta^k_{[i}\state{Y_{j]}}, 
\notag \\
\gen{S}^{ij}\state{\bar{X}^k} & \sim \varepsilon^{ijkm}\state{Y_m}, &  \gen{S}^{ij}\state{X_k} & \sim \delta_k^{[i}\state{\bar{Y}^{j]}}. \label{eq:qaction}
\end{align}
Of course, there is also dependence on the Lorentz indices, but we will not need these precise factors for the full algebra. Simply we note that all interactions allowed by quantum numbers appear with nonzero coefficient. So $\gen{S}^\alpha$ has nonvanishing action only on module elements with $n_{\alpha}>0$.

For $\lambda=0$ the spin-chain states transform in the tensor product of the single-site $\alg{osp}(6|4)$ representations, but beyond leading order they transform in a deformed representation; the $\alg{osp}(6|4)$ generators act on the spin chain via interactions that couple multiple sites.  Beside the manifest $\gen{R}$ and $\gen{L}$ symmetry generators, all generators receive corrections for $\lambda \neq 0$. Perturbatively, these interactions act on an increasing number of sites with each order in $\lambda$. Counting powers of the coupling constant implies that at $\mathcal{O}(\lambda)$, interactions can act on up to a total of four sites (e.g. two initial and two final sites), and up to six sites at $\mathcal{O}(\lambda^2)$. In the planar limit, which is our focus, these interactions act on adjacent modules.

Finally,   $\alg{osp}(6|4)$ has a quadratic Casimir $J^2$, 
\< \label{eq:osp64jsquared}
J^2 \eq \frac{1}{8}\Big(\comm{\gen{Q}_{ij,\alpha}}{\gen{S}^{ij, \alpha}} - 2 \gen{R}^i_j \gen{R}^j_i + 2 \gen{L}^\alpha_\beta\gen{L}^\beta_\alpha + 4 \gen{D}^2 - \acomm{\gen{P}_{\alpha\beta}}{\gen{K}^{\alpha\beta}}\Big).
\>
On highest-weight states, $J^2$ simplifies to
\<
J^2 \eq \half\Big(D(D+3) + s(s+2)  + 3 R^1_1+2 R^2_2 + R^3_3 - \half \sum_{i=1}^4 (R^i_i)^2 \Big)
 \nln
 \eq \half\Big(D(D+3) + s(s+2)  
 \nl
 - \quarter q_1(q_1+2) - \quarter q_2(q_2+2) -\sfrac{1}{8} (2 p + q_1 + q_2)^2 - (2 p + q_1 + q_2)\Big). 
\>
Here $D$ is the dimension and $s$ is the Lorentz spin. The first expression uses eigenvalues of all diagonal entries of the traceless matrix of $\gen{R}$-symmetry generators, while the second uses the standard $\alg{su}(4)$ Dynkin labels,
\[
q_1 = R^2_2-R^1_1, \quad p=R^3_3 - R^2_2, \quad q_2 = R^4_4-R^3_3.
\]
The $\alg{osp}(6|4)$ spin $j$ satisfies $j(j+1) = J^2$, the eigenvalue of $\gen{J}^2$. The tensor product of a conjugate pair of modules $\mathcal{V}_\phi$ and $\mathcal{V}_{\bar \phi}$ has one highest-weight state for each nonnegative integer spin $j$ (and no other highest-weight states),
\[  \label{eq:conjugateproduct}
\mathcal{V}_\phi \otimes \mathcal{V}_{\bar \phi} = \sum_{j=0}^\infty \mathcal{V}_j.
\]
Similarly, a like pair of modules has has one highest-weight state with spin $(j-1/2)$ for each nonnegative integer $j$,
\[ \label{eq:likeproduct}
\mathcal{V}_\phi \otimes \mathcal{V}_\phi  = \sum_{j=0}^\infty \mathcal{V}_{j-1/2}, \quad  \mathcal{V}_{\bar \phi} \otimes \mathcal{V}_{\bar \phi} = \sum_{j=0}^\infty \mathcal{V}_{j-1/2}.
\]

\subsection{Restriction to the $\alg{osp}(4|2)$ sector \label{sec:restrictosp42}}
Consider the set of states that at leading order are annihilated by $\gen{Q}_{12,2}$ and $\gen{S}^{12,2}$, which is a $1/12$ BPS condition\footnote{Beyond leading order, as we will see, generic such states acquire anomalous dimension so that they do not satisfy the exact $1/12$ BPS condition  $R^1{}_1 + R^2{}_2 - L^2{}_2 - D = 0$. Here the full scaling dimension is $D=D_0 + \delta D$, which includes an anomalous contribution.}. According to the last algebra relation of (\ref{eq:anticommutators}),  they satisfy
\[
R^1{}_1 + R^2{}_2 - L^2{}_2 - D_0 = 0, \label{eq:1/12bps}
\]
where ordinary font denotes the eigenvalue of the corresponding generator in Gothic font, and $D_0$ is the classical dimension of a state. Since states only mix with other states with the same Lorentz and $\gen{R}$ symmetry quantum numbers, and we choose a renormalization scheme where only states with the same classical dimension mix, it follows that this set of states is closed to all orders in perturbation theory. 

Under this restriction, $\alg{osp}(6|4)$ reduces to the set of generators that commute with the left side of (\ref{eq:1/12bps}), and from the above algebra relations we find a residual $\alg{u}(1) \ltimes \alg{su}(1|1) \times \alg{osp}(4|2)$ algebra.  Since at leading order the $\alg{su}(1|1)$ algebra (generated by $\gen{Q}_{12,2}$ and $\gen{S}^{12,2}$) acts trivially, we call this sector the $\alg{osp}(4|2)$ sector. $ \alg{osp}(4|2)$ includes a $\alg{sl}(2)$ subalgebra of $\alg{sp}(4)$,  two $\alg{su}(2)$ algebras from the original $\alg{su}(4)$ $\gen{R}$ symmetry, and eight supercharges. More precisely, and introducing a convenient notation, the $\alg{osp}(4|2)$ generators are related to those for the full $\alg{osp}(6|4)$ theory as (all index variables run from 1 to 2)
\begin{align}
 \gen{J}^{11} & =  \half \gen{P}_{11}, 
& 
 \gen{J}^{22} & = \half \gen{K}^{11},
& 
 \gen{J}^{12} & =  -\sfrac{1}{2} \gen{L} + \gen{D} + \half \delta\gen{D},
 \notag \\
\gen{R}^{ab}& =  \half \varepsilon^{ac} \gen{R}^b_c  +\half \varepsilon^{bc} \gen{R}^a_c, 
&
 \tilde{\gen{R}}^{\mathfrak{ab}} & = \half \varepsilon^{\mathfrak{ac}} \gen{R}^{\mathfrak{b}+2}_{\mathfrak{c}+2}  +\half \varepsilon^{\mathfrak{bc}} \gen{R}^{\mathfrak{a}+2}_{\mathfrak{c}+2}, & &
 \notag \\
 \gen{Q}^{a1\mathfrak{b}} & = \varepsilon^{ac}\varepsilon^{\mathfrak{bd}} \gen{Q}_{c(\mathfrak{d}+2),1}, 
&
 \gen{Q}^{a2\mathfrak{b}} & = - \gen{S}^{a(\mathfrak{b}+2),1}. & & \label{eq:osp42generators}
\end{align}
Here we have also introduced $\gen{L}$, which generates an additional $\alg{u}(1)$ and is related to $\alg{osp}(6|4)$ generators as
\[
\gen{L} = \gen{R}^1_1 + \gen{R}^2_2 = -\gen{L}^1_1 + \gen{D}_0, \label{eq:definel}
\]
The second equality is satisfied within this sector due to (\ref{eq:1/12bps}). $\gen{L}$ commutes with the $\alg{osp}(4|2)$ generators. As we will see below, due to the restricted field content in this sector $\gen{L}$ simply gives half of the number of spin-chain sites (the number of pairs of conjugate representations). 

The rank-one subalgebras take the standard form
\[
\comm{J^{AB}}{J^{CD}} = \varepsilon^{CB}J^{AD} - \varepsilon^{AD}J^{CB}, \quad \comm{J^{AB}}{X^{C}} = \half \varepsilon^{CB}X^{A} + \half  \varepsilon^{CA}X^{B}, \label{eq:rankonealgebras}
\]
and the anticommutators are
\[
\acomm{\gen{Q}^{a\beta\mathfrak{c}}}{\gen{Q}^{d\epsilon\mathfrak{f}}} = -\varepsilon^{\beta\epsilon}\varepsilon^{\mathfrak{cf}} \gen{R}^{ad} -\varepsilon^{ad}\varepsilon^{\beta\epsilon} \tilde{\gen{R}}^{\mathfrak{cf}} + 2 \varepsilon^{ad}\varepsilon^{\mathfrak{cf}} \gen{J}^{\beta\epsilon}. \label{eq:osp42anticom}
\]

We use hatted notation for the $\alg{su}(1|1)$ algebra supercharges  (note the reversed order of $\alg{R}$ indices for the second one),
\[
\lambda \hat{\gen{Q}} = \gen{Q}_{12,2}, \quad \lambda \hat{\gen{S}} = \gen{S}^{21,2}.
\]
In contrast to the introduction, we have now included a  factor of $\lambda$ in the definitions since these generators act nontrivially first at $\mathcal{O}(\lambda)$ in this sector. $\hat{\gen{Q}}$ and $\hat{\gen{S}}$ are nilpotent and the only nonvanishing (anti)commutator for $\alg{su}(1|1)$ is 
\[ \label{eq:qhatshatacomm}
\acomm{\hat{\gen{Q}}}{\hat{\gen{S}}} =  \frac{1}{\lambda^2} \delta \gen{D} =  \mathcal{H}.
\]
Here we have introduced $\mathcal{H}$, the anomalous part of the dilatation generator divided by $\lambda^2$. We will also normalize the (one-loop) supercharges so that 
\[
\hat{\gen{S}} = (\hat{\gen{Q}})^\dagger.
\]
We can do this since only the product of their overall normalizations appears in the Hamiltonian.

Again, $\alg{su}(1|1)$ and $\alg{osp}(4|2)$ commute, and therefore $\delta \gen{D}$ is a shared central charge.  The $\alg{u}(1)$ length generator $\gen{L}$ satisfies
\[ \label{eq:lcomm}
\comm{\gen{L}}{\hat{\gen{Q}}} = \hat{\gen{Q}}, \quad \comm{\gen{L}}{\hat{\gen{S}}} = -\hat{\gen{S}}.
\]
In this work we will only consider the leading (nonvanishing) contributions to the generators. So after this section, $\gen{Q}$, $\gen{J}$, $\hat{\gen{Q}}$, $\hat{\gen{S}}$ and $\mathcal{H}$  will refer only to $\mathcal{O}(\lambda^0)$ terms. We will still explicitly refer to the order of the leading anomalous piece of the dilatation generator, $\delta \gen{D}_2$. 

\subsection{The $\alg{osp}(4|2)$ module and leading-order representation}
The two modules of the $\alg{osp}(4|2)$ sector $\mathcal{V}_\phi^{(4|2)}$ and $\mathcal{V}_{\bar \phi}^{(4|2)}$ can be obtained by acting with the generators (\ref{eq:osp42generators}) on the highest weight states $\state{\phi_1^{(0,0)}}$ and $\state{(\bar{\phi}^4)^{(0,0)}}$. From (\ref{eq:rsymaction} - \ref{eq:qaction}), we conclude states in this sector have  lower $\gen{R}$ indices $1$ or $2$ or upper indices $3$ or $4$.  Then, using  (\ref{eq:lorentanddilatationaction}), we see that (\ref{eq:1/12bps}) implies that all states in this sector have only nonzero values for the first Lorentz excitation number, so we can use a single argument for the Lorentz indices of states as
\[
\state{\phi_a^{(n)}} = \state{\phi_a^{(2n, 0)}} \sim \mathcal{D}_{11}^{n} \phi_a, \quad  \state{\psi_a^{(n)}} = \state{\psi_a^{(2n+1, 0)}} \sim \mathcal{D}_{11}^{n} \psi_{1a}, \quad a=1,2.
\]
Again, the conjugate fields have upper $\gen{R}$ indices $3$, $4$ in the notation for the full theory; it is convenient to replace them with lower indices $\mathfrak{a,b}=1,2$, resulting in 
\<
\state{\bar{\phi}_\mathfrak{a}^{(n)}} \eq \varepsilon_{\mathfrak{ab}} \state{(\bar{\phi}^{\mathfrak{b}+2})^{(2n, 0)}}  \sim \varepsilon_{\mathfrak{ab}} \mathcal{D}_{11}^{n}\bar{\phi}^{\mathfrak{b}+2}, 
\nln
 \state{\bar{\psi}_\mathfrak{a}^{(n)}} \eq \varepsilon_{\mathfrak{ab}} \state{(\bar{\psi}^{\mathfrak{b}+2})^{(2n+1, 0)}} \sim \varepsilon_{\mathfrak{ab}} \mathcal{D}_{11}^{n}\bar{\psi}^{\mathfrak{b}+2}_1.
\>
(\ref{eq:definel}) and (\ref{eq:lorentanddilatationaction}) now imply in this sector
that any state $\state{X}$ satisfies
\[
\gen{L} \state{X} = \half \state{X}. 
\]
In other words, $\gen{L}$ gives $L$ on an $\alg{osp}(4|2)$ spin chain state of  $2 L$ sites, as mentioned previously.  

In the just-introduced notation, (\ref{eq:rsymaction}) implies that the $\alg{su}(2)$ subalgebras' generators $\gen{R}$ act canonically on the the unbarred states, and the $\tilde{\gen{R}}$ act the same way on barred states,
\[
\gen{R}^{ab}\state{X_c} = \half \delta^a_c \varepsilon^{bd} \state{X_d} +  \half \delta^b_c \varepsilon^{ad} \state{X_d}, \quad \tilde{\gen{R}}^{\mathfrak{ab}}\state{\bar{X}_{\mathfrak{c}}} = \half \delta^{\mathfrak{a}}_{\mathfrak{c}} \varepsilon^{\mathfrak{bd}} \state{\bar{X}_{\mathfrak{d}}} +  \half \delta^{\mathfrak{b}}_{\mathfrak{c}} \varepsilon^{\mathfrak{ad}} \state{\bar{X}_{\mathfrak{d}}}.
\]
Closure of the algebra (\ref{eq:rankonealgebras} - \ref{eq:osp42anticom}) fixes the supercharges' action (up to physically irrelevant possible changes of basis) as 
\begin{align}
\gen{Q}^{a1\mathfrak{b}} \state{\phi_c^{(n)}}  & = \delta^a_c \varepsilon^{\mathfrak{bd}} \sqrt{2n + 1} \state{\bar{\psi}^{(n)}_{\mathfrak{d}}}, & \gen{Q}^{a1\mathfrak{b}} \state{\psi_c^{(n)}}  & = \delta^a_c \varepsilon^{\mathfrak{bd}} \sqrt{2n + 2} \state{\bar{\phi}^{(n+1)}_{\mathfrak{d}}},
\notag \\
\gen{Q}^{a2\mathfrak{b}} \state{\phi_c^{(n)}}  & = \delta^a_c \varepsilon^{\mathfrak{bd}} \sqrt{2n } \state{\bar{\psi}^{(n-1)}_{\mathfrak{d}}}, & \gen{Q}^{a2\mathfrak{b}} \state{\psi_c^{(n)}}  & = \delta^a_c \varepsilon^{\mathfrak{bd}} \sqrt{2n + 1} \state{\bar{\phi}^{(n)}_{\mathfrak{d}}}
\notag \\
\gen{Q}^{a1\mathfrak{b}} \state{\bar{\phi}_{\mathfrak{c}}^{(n)}}  & = - \delta^\mathfrak{b}_\mathfrak{c} \varepsilon^{ad} \sqrt{2n + 1} \state{\psi^{(n)}_{d}}, & \gen{Q}^{a1\mathfrak{b}} \state{\bar{\psi}_{\mathfrak{c}}^{(n)}}  & = - \delta^\mathfrak{b}_\mathfrak{c} \varepsilon^{ad} \sqrt{2n + 2} \state{\phi^{(n+1)}_{d}},
\notag \\
\gen{Q}^{a2\mathfrak{b}} \state{\bar{\phi}_{\mathfrak{c}}^{(n)}}  & = - \delta^\mathfrak{b}_\mathfrak{c} \varepsilon^{ad} \sqrt{2n } \state{\psi^{(n-1)}_{d}}, & \gen{Q}^{a2\mathfrak{b}} \state{\bar{\psi}_{\mathfrak{c}}^{(n)}}  & = - \delta^\mathfrak{b}_\mathfrak{c} \varepsilon^{ad} \sqrt{2n + 1} \state{\phi^{(n)}_{d}}.
\end{align}
Note the symmetry between the supercharge actions on barred and unbarred states, only differing by a minus sign and appropriate interchanges of Gothic and Latin indices.  
Finally, the action of the $\gen{J}$ (independent of $\gen{R}$ and $\tilde{\gen{R}}$ indices) is the same for barred and unbarred states.  Representing both $\phi_a$ and $\bar{\phi}_{\mathfrak{a}}$ with $\phi$,  and   $\psi_a$ and $\bar{\psi}_{\mathfrak{a}}$ with $\psi$, we have
\begin{align}
\gen{J}^{11} \state{\phi^{(n)}} & =  \sqrt{(n+\half)(n+1)} \state{\phi^{(n+1)}}, & \gen{J}^{11} \state{\psi^{(n)}} & = \sqrt{(n+1)(n+\sfrac{3}{2})} \state{\psi^{(n+1)}},
\notag \\
\gen{J}^{22} \state{\phi^{(n)}} & = \sqrt{(n-\half)n} \state{\phi^{(n-1)}}, & \gen{J}^{22} \state{\psi^{(n)}} & =  \sqrt{n(n+\half)} \state{\psi^{(n-1)}},
\notag \\
\gen{J}^{12} \state{\phi^{(n)}} & = (n +\sfrac{1}{4}) \state{\phi^{(n)}}, & \gen{J}^{12} \state{\psi^{(n)}} & = (n +\sfrac{3}{4}) \state{\psi^{(n)}}.
\end{align}
In the basis we have chosen, Hermiticity is manifest,
\[
(\gen{Q}^{a1 \mathfrak{b}})^\dagger = \varepsilon_{ac}\varepsilon_{\mathfrak{bd}}\gen{Q}^{c2\mathfrak{d}}, \quad (\gen{J}^{11})^\dagger = \gen{J}^{22}, \quad (\gen{J}^{12})^\dagger = \gen{J}^{12},  \quad (\gen{X}^{AB})^\dagger = \varepsilon_{AD}\varepsilon_{CB} \gen{X}^{CD},
\]
where $\gen{X}$ denotes $\gen{R}$ or $\tilde{\gen{R}}$.  

The quadratic Casimir for this sector is %
\[
\gen{J}^2=-\quarter \varepsilon_{ad}\varepsilon_{cb} \gen{R}^{ab}\gen{R}^{cd} -\quarter \varepsilon_{\mathfrak{ad}}\varepsilon_{\mathfrak{cb}} \tilde{\gen{R}}^{\mathfrak{ab}}{\gen{R}}^{\mathfrak{cd}} + \half  \varepsilon_{\alpha \delta}\varepsilon_{\gamma \beta} \gen{J}^{\alpha \beta} \gen{J}^{\gamma \delta} -\quarter \varepsilon_{ad}\varepsilon_{\beta \epsilon } \varepsilon_{\mathfrak{cd}} \gen{Q}^{a\beta\mathfrak{c}}\gen{Q}^{d\epsilon\mathfrak{d}}.  \label{eq:defineosp42casimir}
\]
It follows from the algebra commutation relations that highest-weight states, which are annihilated by  $\gen{R}^{22}$, $\tilde{\gen{R}}^{22}$, $\gen{J}^{22}$ and $\gen{Q}^{a2\mathfrak{b}}$,  have quadratic Casimir eigenvalue
\[ \label{eq:twositeCartan}
J^2 = - \half R^{12}(R^{12}-1)- \half \tilde{R}^{12}(\tilde{R}^{12} - 1)   + J^{12}(J^{12} + 1).
\]
As for $\alg{osp}(6|4)$, a eigenstate of the quadratic Casimir has $\alg{osp}(4|2)$ spin $j$ given by  $J^2 = j(j+1)$.  

It is a straightforward exercise to find the highest-weight states for two-site states. For a conjugate pair of modules, there is one highest-weight state (and irreducible highest-weight $\alg{osp}(4|2)$ module) for each nonnegative integer spin $j$, matching precisely the result previously given for the full $\alg{osp}(6|4)$ modules (\ref{eq:conjugateproduct}),
\[ \label{eq:42conjugateproduct}
\mathcal{V}_\phi^{(4|2)} \otimes \mathcal{V}_{\bar \phi}^{(4|2)} = \sum_{j=0}^\infty \mathcal{V}^{(4|2)}_j.
\]
 The corresponding Cartan charges are  
\[ \label{eq:twositedifferenthigestweight}
[ R^{12}, J^{12}, \tilde{R}^{12} ] = [-\half, \half, -\half] \quad \text{and} \quad [ R^{12}, J^{12}, \tilde{R}^{12} ] = [0, j, 0] \quad  j =1, 2, \ldots 
\]
We will also need the case of two identical modules (which would appear as two next-nearest neighbor sites on the alternating chain). Again there is precise agreement with the $\alg{osp}(6|4)$ result (\ref{eq:likeproduct}), 
\[ \label{eq:42likeproduct}
\mathcal{V}_\phi^{(4|2)} \otimes \mathcal{V}^{(4|2)}_\phi  = \sum_{j=0}^\infty \mathcal{V}^{(4|2)}_{j-1/2}, \quad  \mathcal{V}^{(4|2)}_{\bar \phi} \otimes \mathcal{V}^{(4|2)}_{\bar \phi} = \sum_{j=0}^\infty \mathcal{V}^{(4|2)}_{j-1/2}.
\]
 and the $\alg{osp}(4|2)$ Cartan charges are
\[ \label{eq:twositeidenticalhigestweight}
[-1,  \half, 0] \quad  \text{and} \quad  [0, j + \half, 0], \quad j = 0, 1, \ldots
\]

Finally, the $\alg{osp}(4|2)$ subsector has  $\alg{sl}(2)$ subsector(s) in which only the $\gen{J}$ act nontrivially\footnote{Of course, $\gen{L}$ also still acts within this sector, and is proportional to the action of $\gen{R}^{12}$.}. The two modules are spanned by 
\[
\state{\phi_1^{(n)}} \quad \text{and} \quad \state{\psi_1^{(n)}}. 
\]
The $\alg{sl}(2)$ sector two-site highest-weight states (which are descendants in the $\alg{osp}(4|2)$ sector) have $\alg{sl}(2)$ spins that take the same value as for the larger $\alg{osp}(4|2)$ sector, and therefore as in $\alg{osp}(6|4)$ as well.

\subsection{Light-cone superspace basis}

We will find it very useful to use a light-cone superspace \cite{Kogut:1969xa,Brink:1982pd,Mandelstam:1982cb,Belitsky:2004yg}  basis for the modules\footnote{I thank A. Belitsky for suggesting such a basis, actually in the context of $\mathcal{N}=4$ SYM.}.  This basis parameterizes $\mathcal{V}_{\phi}^{(4|2)}$ ($\mathcal{V}_{\bar \phi}^{(4|2)}$) with continuous variables: $x$, a $\alg{su}(2)$ doublet $\theta^a$ ($\bar{\theta}^{\mathfrak{a}}$), and an anticommuting $\alg{su}(2)$ doublet $\bar{\eta}^{\mathfrak{a}}$ ($\eta^a$). The spin-chain states are labeled $\state{x, \theta, \bar{\eta}}$, or  $\state{x, \bar{\theta}, \eta}$, which are defined via sums over the entire modules,

\< \label{eq:definelightcone}
\state{x, \theta, \bar{\eta}} \eq \sum_{n=0}^\infty \frac{x^n \theta^a}{n!} \sqrt{(2n)!} \state{\phi_a^{(n)}}  + \frac{x^n \bar{\eta}^{\mathfrak{a}}}{n!} \sqrt{(2n+1)!} \state{\bar{\psi}_{\mathfrak{a}}^{(n)}}, 
\nln
\state{x, \bar{\theta}, \eta} \eq \sum_{n=0}^\infty \frac{x^n \bar{\theta}^{\mathfrak{a}}}{n!}\sqrt{(2n)!}  \state{\bar{\phi}_{\mathfrak{a}}^{(n)}}  + \frac{x^n \eta^a}{n!} \sqrt{(2n+1)!} \state{\psi_a^{(n)}}.
\>
In this basis the leading-order generators are represented by differential operators. As an example, we derive the representation for $\gen{Q}^{a1\mathfrak{c}}$ acting on $\mathcal{V}_{\phi}^{(4|2)}$, 
\<
\gen{Q}^{a1\mathfrak{c}} \state{x, \theta, \bar{\eta}} \eq \sum_{n=0}^{\infty} \frac{x^n}{n!} \Big(\sqrt{(2n)!} \theta^b \gen{Q}^{a1\mathfrak{c}} \state{\phi_b^{(n)}} - \sqrt{(2n+1)!} \bar{\eta}^{\mathfrak{d}}  \gen{Q}^{a1\mathfrak{c}} \state{\psi_{\mathfrak{d}}^{(n)}} \Big)
\nln
\eq 
\sum_{n=0}^{\infty} \frac{x^n}{n!} \Big(\sqrt{(2n+1)!} \theta^a \varepsilon^{\mathfrak{cd}} \state{\bar{\psi}_{\mathfrak{d}}^{(n)}} +  \sqrt{(2n+2)!} \varepsilon^{ab} \bar{\eta}^{\mathfrak{c}}  \state{\phi_b^{(n+1)}} \Big)
\nln
\eq \Big(\theta^a \bar{\partial}^{\mathfrak{c}} + \partial_x \partial^a \bar{\eta}^{\mathfrak{c}}   \Big)  \state{x, \theta, \bar{\eta}} .
\>
Our conventions for raising and lowering indices lead to, for $\mathcal{V}_{\phi}^{(4|2)}$,
\begin{align}
\partial^{a} & = \varepsilon^{ab} \partial_{\theta^{b}} = \varepsilon^{ab} \partial_b, & \bar{\partial}^{\mathfrak{c}}&  = \varepsilon^{\mathfrak{cd}} \partial_{\bar{\eta}^{\mathfrak{d}}} = \varepsilon^{\mathfrak{cd}} \bar{\partial}_{\mathfrak{d}}, \notag \\ 
\partial_{a} & =   \partial^b \varepsilon_{ba}, & \partial_{\mathfrak{c}}&  =  \bar{\partial}^{\mathfrak{d}} \varepsilon_{\mathfrak{dc}} .
\end{align}
For $\mathcal{V}_{\bar \phi}^{(4|2)}$, the same equations apply for $\theta$ replaced by $\eta$ and $\bar{\eta}$ replaced by $\bar{\theta}$. So, for example, $\partial^a$ is bosonic acting on $\mathcal{V}_{\phi}^{(4|2)}$, and fermionic on $\mathcal{V}_{\bar \phi}^{(4|2)}$. 
We repeat the above result for $\gen{Q}^{a1\mathfrak{c}}$ and add the parallel expressions for the other supercharges and for $\mathcal{V}_{\bar \phi}^{(4|2)}$ (abbreviating the states by suppressing an $\eta$ or $\bar{\eta}$),
\begin{align}
\gen{Q}^{a1\mathfrak{c}} \state{x, \theta} &=\Big(\theta^a \bar{\partial}^{\mathfrak{c}} + \partial_x \partial^a \bar{\eta}^{\mathfrak{c}}   \Big)  \state{x, \theta} , &   \gen{Q}^{a2\mathfrak{c}} \state{x, \theta} &=\Big(2 x \theta^a \bar{\partial}^{\mathfrak{c}} + (2 x \partial_x +1)\partial^a \bar{\eta}^{\mathfrak{c}}   \Big)  \state{x, \theta} ,
\notag \\
\gen{Q}^{a1\mathfrak{c}} \state{x, \bar{\theta}} &=-\Big(\bar{\theta}^{\mathfrak{c}} \partial^{a} + \partial_x \bar{\partial}^{\mathfrak{c}} \eta^a   \Big)  \state{x, \bar{\theta}} , &  \gen{Q}^{a2\mathfrak{c}} \state{x, \bar{\theta}} &=-\Big(2 x \bar{\theta}^{\mathfrak{c}} \partial^{a} + (2 x \partial_x +1) \bar{\partial}^{\mathfrak{c}} \eta^a   \Big)  \state{x, \bar{\theta}}.
\end{align}
The remaining generators' actions can be written the same way\footnote{Provided we project the right sides to their bosonic components. For example, the first term for $\gen{R}^{ab}$ only includes $\theta^{\{a}\varepsilon^{b\}c} \partial_{\theta^c}$ and \emph{not} $\theta^{\{a}\varepsilon^{b\}c} \partial_{\eta^c}$.} for both types of modules,  
\begin{align} \label{eq:osp42lightcone}
\gen{R}^{ab} &= \theta^{\{a} \partial^{b\}} + \eta^{\{a} \partial^{b\}}, & \bar{\gen{R}}^{\mathfrak{cd}} &= \bar{\theta}^{\{\mathfrak{c}} \partial^{\mathfrak{d}\}} + \bar{\eta}^{\{\mathfrak{c}} \partial^{\mathfrak{d}\}},
\notag \\
\gen{J}^{11} & = \partial_x, & \gen{J}^{22} & = 2 x^2 \partial_x + x - 2 x \Big(\varepsilon_{ab} \eta^a \partial^b + \varepsilon_{\mathfrak{cd}} \bar{\eta}^c \bar{\partial}^{\mathfrak{d}}\Big), 
\notag \\
 \gen{J}^{12} & = x \partial_x + \frac{1}{4} -\frac{1}{2} \Big(\varepsilon_{ab} \eta^a \partial^b + \varepsilon_{\mathfrak{cd}} \bar{\eta}^c \bar{\partial}^{\mathfrak{d}}\Big). & &
\end{align}
%

\section{Leading corrections in the  $\alg{osp}(4|2)$ sector \label{sec:oneloop}}
We will now compute the leading actions of $\hat{\gen{Q}}$ and $\hat{\gen{S}}$, which immediately give the two-loop dilatation generator. 

\subsection{Structure of $\hat{\gen{Q}}$ and $\hat{\gen{S}}$ interactions}
(\ref{eq:lcomm}) implies that $\hat{\gen{Q}}$ and $\hat{\gen{S}}$ are dynamic; they change the length of the spin chain as do spin-chain generators in $\mathcal{N}=4$ SYM \cite{Beisert:2003ys}. $\hat{\gen{Q}}$ inserts two sites, and $\hat{\gen{S}}$ removes two sites. Note that this means these supercharges have well-defined actions only on cyclic states, which is all that is required since these spin-chain states represent single-trace local operators.  At one-loop, consistency with the coupling constant dependence of the interactions of the Lagrangian requires $\hat{\gen{Q}}$ to replace one site with three, and $\hat{\gen{S}}$ to replace three sites with one. Let $\mathcal{U}$  be the generator that shifts all sites by two to the right (with the last site $2 L$ going to site $2$ for example). By definition a cyclic alternating state $\state{Y}$ satisfies $\mathcal{U}\state{Y}=\state{Y}$. $\hat{\gen{Q}}$ then acts as 
\[
\hat{\gen{Q}} \state{Y} = \frac{L}{L +1} \sum_{i=0}^{L} \mathcal{U}^{-i} \Big(\hat{\gen{Q}}(1)  + \hat{\gen{Q}}(2)\Big)\state{Y} .
\]
and $\hat{\gen{Q}}(i)$ gives the action of $\hat{\gen{Q}}$ the $i$th site, which we determine below. Similarly
\[
\hat{\gen{S}} \state{Y} = \frac{L}{L -1} \sum_{i=0}^{L-2} \mathcal{U}^{-i} \Big(\hat{\gen{S}}(1,2,3)  + \hat{\gen{S}}(2, 3, 4)\Big)\state{Y} .
\]
In both these expressions a minus signs must be included for each crossing of two fermions (or for a supercharge crossing a fermion).

Constraints from manifest $\gen{R}$ and $\tilde{\gen{R}}$ symmetries and consistency with the classical scaling dimension assignments further severely restrict the supercharge actions.  For instance, acting on a  scalar initial state, we (almost) immediately can restrict to an ansatz for the action of $\hat{\gen{Q}}$ on one site,
\< \label{eq:qhatansatz}
\hat{\gen{Q}}\state{\phi_a^{(m)}} \eq \sum_{n+p < m} \Bigg( c_1(m, n, p) \varepsilon^{bc} \state{\phi_a^{(n)} \psi_b^{(p)} \phi_c^{(m-n-p-1)}} 
\nl
- c_1'(m, n, p) \varepsilon^{bc} \state{\phi_c^{(m-n-p-1)} \psi_b^{(p)} \phi_a^{(n)}} \Big)
\nl
+  \varepsilon^{\mathfrak{bc}} \Big(c_2(m, n, p)  \state{\phi_a^{(n)} \bar{\phi}_{\mathfrak{b}}^{(p)} \bar{\psi}_{\mathfrak{c}}^{(m-n-p-1)}} - c_2'(m, n, p) \state{\bar{\psi}_{\mathfrak{c}}^{(m-n-p-1)} \bar{\phi}_{\mathfrak{b}}^{(p)}  \phi_a^{(n)}}\Big).
\nl
\>
Here and below summations will be over all nonnegative integers satisfying an inequality. In this case we have $n=0, 1, \ldots m-1$ and $p = 0, 1, \ldots m-n-1$. This ansatz (\ref{eq:qhatansatz}) is the most general one  consistent with the constraints mentioned above and two further observations.  First, note the $\alg{su}(2)$ identity,
\[
 \varepsilon^{bc} \state{X_bY_cZ_a} + \varepsilon^{bc} \state{X_aY_bZ_c}=  \varepsilon^{bc} \state{X_bY_aZ_c},
 \]
which eliminates the need for another term. Second, a priori a scalar to three fermions interaction would be allowed, but this turns out to be inconsistent with the requirement that $\hat{\gen{Q}}$ commutes with all of the $\alg{osp}(4|2)$ generators (one can check this using the same methods we will now use to determine the $c_i$). 

\subsection{Constraints from commutator with $\gen{J}^{11}$}
Rather than considering the commutators with all $\alg{osp}(4|2)$ generators, let us first just consider 
\[
\comm{\hat{\gen{Q}}}{\gen{J}^{11}}=0 \text{ on cyclic states.}
\]
As initially noted for $\mathcal{N}=4$ SYM \cite{Beisert:2003ys}, the fact that the algebra only must be satisfied on cyclic spin chain states allows for  ``gauge transformations''. Here, consistency with the basic constraints used above and with alternating modules restricts such a gauge transformation to
\< \label{eq:gauge}
\comm{\hat{\gen{Q}}}{\gen{J}^{11}}\state{X} \eq (-1)^X g_1 \varepsilon^{\mathfrak{ab}} \state{X\bar{\phi}_{\mathfrak{a}}^{(0)}\bar{\psi}_{\mathfrak{b}}^{(0)}} - g_2 \varepsilon^{\mathfrak{ab}} \state{\bar{\psi}_{\mathfrak{a}}^{(0)}\bar{\phi}_{\mathfrak{b}}^{(0)}X}
\nl
+ (-1)^X g_3 \varepsilon^{ab}\state{X\psi_a^{(0)}\phi_b^{(0)}} - g_4 \varepsilon^{ab}\state{\phi_a^{(0)}\psi_b^{(0)}X},
\nln
\comm{\hat{\gen{Q}}}{\gen{J}^{11}}\state{\bar{X}} \eq (-1)^{\bar{X}} g_2 \varepsilon^{\mathfrak{ab}} \state{\bar{X}\bar{\psi}_{\mathfrak{a}}^{(0)}\bar{\phi}_{\mathfrak{b}}^{(0)}} - g_1 \varepsilon^{\mathfrak{ab}} \state{\bar{\phi}_{\mathfrak{a}}^{(0)}\bar{\psi}_{\mathfrak{b}}^{(0)}\bar{X}}
\nl
+ (-1)^{\bar{X}} g_4 \varepsilon^{ab}\state{\bar{X}\phi_a^{(0)}\psi_b^{(0)}} - g_3 \varepsilon^{ab}\state{\psi_a^{(0)}\phi_b^{(0)}\bar{X}}.
\>
$X$ represents any element of $\mathcal{V}_{\phi}^{(4|2)}$, $\bar{X}$ represents any element in $\mathcal{V}_{\bar \phi}^{(4|2)}$, and $(-1)^X$ gives $(-1)$ for fermionic $X$ and $1$ otherwise. To see that this interaction gives zero on cyclic states, consider the two terms with coefficient $g_1$. The first $g_1$ term  inserts an $\tilde{\gen{R}}$ singlet to the right of $\mathcal{V}_{\phi}^{(4|2)}$ sites. But on alternating cyclic (or infinite) chains,  this is canceled by  the second $g_1$ term, which inserts the same $\tilde{\gen{R}}$ singlet to the left of $\mathcal{V}_{\bar \phi}^{(4|2)}$ sites with a relative minus sign. The cancellations for the other $g_i$ terms work in the same way.

Since $\gen{J}^{11}$ does not affect $\gen{R}$ charges, we obtain independent equations for the four coefficient functions appearing in (\ref{eq:qhatansatz}). For instance the unprimed $c_1$ terms contribute
\< \label{eq:j11constraint}
\comm{\hat{\gen{Q}}}{\gen{J}^{11}}_{c_1} \state{\phi_a^{(m)}} \eq  \sum_{n+p \leq m} \bigg( \sqrt{(m+1)(m+\half)} c_1(m+1, n, p) 
\nl
- \sqrt{n(n-\half)} c_1(m, n-1, p) - \sqrt{ p ( p +\half)}c_1(m, n, p-1)
\nl
 -\sqrt{(m-n-p)(m-n-p-\half)}c_1(m, n, p)\bigg)\varepsilon^{bc} \state{\phi_a^{(n)}\psi_b^{(p)}\phi_c^{(m-n-p)}}
 \nln
 \eq g_3 \varepsilon^{bc} \state{\phi_a^{(0)} \psi_b^{(0)} \phi_c^{(m)}}.
\>
From acting on a scalar with $m=0$, we find the only coefficient with first argument $1$,  $c_1(1, 0,0) = \sqrt{2} g_3$. But now it is straightforward to see that all coefficients are determined inductively. Assume all coefficients with first argument less than or equal to $m_0$ are known. Then (\ref{eq:j11constraint}) determines all coefficients with first argument $m_0+1$ in terms of these known coefficients and $g_3$.  The solution for all arguments is\footnote{We could also write $r_{\pm}(x) = \sqrt{\frac{\sqrt{\pi/2}\Gamma(x+1)}{2^{x\pm \half}\Gamma(x+1\pm\half)}} \rightarrow \sqrt{\frac{\sqrt{\pi}\Gamma(x+1)}{\Gamma(x+1\pm\half)}}$. Up to a factor of $\sqrt{2}$ in one case, we can use the second expression since the powers of $2$ will cancel in the expressions for the $c_i$ given here and later.} 
\[
 c_1(m,n, p) =  \frac{\sqrt{2} g_3r_-(m)}{(m-n)r_-(n)r_+(p)r_-(m-n-p-1)}, \quad r_{\pm}(x) = \frac{\sqrt{x!}}{\sqrt{(2 x \pm 1)!!}}   .
\]

Repeating for the other three coefficient functions entering (\ref{eq:qhatansatz}), again yields solutions determined inductively from a single coefficient. These solutions depend in total on the four $g_i$ coefficients appearing in (\ref{eq:gauge}). Furthermore, (\ref{eq:gauge}) implies that these are the only free parameters for $\hat{\gen{Q}}$  for acting on scalars in $\mathcal{V}_{\bar \phi}^{(4|2)}$. For $\hat{\gen{Q}}$ acting on fermions, in addition to structures paralleling the four terms of (\ref{eq:qhatansatz}), there is an additional fermion-to-three-scalar interaction that is possible, which commutes exactly with $\gen{J}^{11}$. Therefore, the $\gen{J}^{11}$ constraint allows for two more free parameter (fermions and conjugate fermions), in addition to the four $g_i$. However, there are more constraints.

\subsection{Anticommutator with $\gen{Q}^{a1\mathfrak{b}}$ and solution for $\hat{\gen{Q}}$ and $\hat{\gen{S}}$}
Of course the $\alg{osp}(4|2)$ supercharges relate coefficients unrelated by the $\gen{J}^{11}$.  Then requiring the anticommutator with $\gen{Q}^{a1\mathfrak{b}}$ to vanish on cyclic states fixes all of the above independent coefficients to be proportional to a single free parameter; $\hat{\gen{Q}}$ is determined by symmetry up to overall normalization. This free parameter is fixed by any single nonvanishing two-loop anomalous dimension, since $\hat{\gen{S}} =  (\hat{\gen{Q}})^\dagger$, and  (\ref{eq:qhatshatacomm}) $\acomm{\hat{\gen{Q}}}{\hat{\gen{S}}} =  \delta \gen{D}_2$. The correct choice turns out to be $g_3=1/2$, as we will see below.  With this normalization, the gauge transformation for the anticommutator between supercharges is
\< \label{eq:qqhatgauge}
\acomm{\hat{\gen{Q}}}{\gen{Q}^{a1\mathfrak{b}}}\state{X} \eq \varepsilon^{ac} \varepsilon^{\mathfrak{bd}}\Big(\state{\phi_c^{(0)}\bar{\phi}^{(0)}_\mathfrak{d}X}- \state{X\bar{\phi}^{(0)}_\mathfrak{d}\phi_c^{(0)}}  \Big) ,
\nln
\acomm{\hat{\gen{Q}}}{\gen{Q}^{a1\mathfrak{b}}}\state{\bar{X}} \eq \varepsilon^{ac} \varepsilon^{\mathfrak{bd}}\Big( \state{\bar{\phi}^{(0)}_\mathfrak{d}\phi_c^{(0)}\bar{X}} -\state{\bar{X}\phi_c^{(0)}\bar{\phi}^{(0)}_\mathfrak{d}} \Big).
\>

The corresponding complete solution for $\hat{\gen{Q}}$ acting on $\mathcal{V}_{\phi}^{(4|2)}$ depends on three coefficient functions $c_i$ (note that $c_1$ is as previously, but $c_2$ is different from the $c_2$ appearing in (\ref{eq:qhatansatz})),
\< \label{eq:qhatsolution}
\hat{\gen{Q}}\state{\phi_a^{(m)}} \eq \sum_{n+p < m}  c_1(m, n, p) \varepsilon^{bc} \Big(   \state{\phi_a^{(n)} \psi_b^{(p)} \phi_c^{(m-n-p-1)}}  -  \state{\phi_c^{(m-n-p-1)} \psi_b^{(p)} \phi_a^{(n)}} \Big)
\nl
+   c_1(m, n, m-n-p-1) \varepsilon^{\mathfrak{bc}} \Big( \state{\phi_a^{(n)} \bar{\phi}_{\mathfrak{b}}^{(p)} \bar{\psi}_{\mathfrak{c}}^{(m-n-p-1)}} -  \state{\bar{\psi}_{\mathfrak{c}}^{(m-n-p-1)} \bar{\phi}_{\mathfrak{b}}^{(p)}  \phi_a^{(n)}}\Big),
\nln
\hat{\gen{Q}}\state{\bar{\psi}_{\mathfrak{a}}^{(m)}} \eq \sum_{n+p < m} c_2(m, n, p) \varepsilon^{bc} \Big(   \state{\bar{\psi}_{\mathfrak{a}}^{(n)} \psi_b^{(p)} \phi_c^{(m-n-p-1)}}  +  \state{\phi_c^{(m-n-p-1)} \psi_b^{(p)} \bar{\psi}_{\mathfrak{a}}^{(n)}} \Big)
\nl
+   c_2(m, n, m-n-p-1) \varepsilon^{\mathfrak{bc}} \Big( \state{\bar{\psi}_{\mathfrak{a}}^{(n)} \bar{\phi}_{\mathfrak{b}}^{(p)} \bar{\psi}_{\mathfrak{c}}^{(m-n-p-1)}} +  \state{\bar{\psi}_{\mathfrak{c}}^{(m-n-p-1)} \bar{\phi}_{\mathfrak{b}}^{(p)}  \bar{\psi}_{\mathfrak{a}}^{(n)}}\Big)
\nl
+ \sum_{n+p \leq m} c_3(m, n, p) \varepsilon^{bc} \state{\phi_b^{(n)}\bar{\phi}_{\mathfrak{a}}^{(p)}\phi_c^{(m-n-p)}}.
\>
The action on $\mathcal{V}_{\bar \phi}^{(4|2)}$ simply involves switching barred and unbarred module elements and an overall minus sign,
\< \label{eq:qhatsolutionconjugate}
\hat{\gen{Q}}\state{\bar{\phi}_{\mathfrak{a}}^{(m)}} \eq -\sum_{n+p < m}  c_1(m, n, p)\varepsilon^{\mathfrak{bc}}\Big(   \state{\bar{\phi}_{\mathfrak{a}}^{(n)} \bar{\psi}_{\mathfrak{b}}^{(p)} \bar{\phi}_{\mathfrak{c}}^{(m-n-p-1)}}  -  \state{\bar{\phi}_{\mathfrak{c}}^{(m-n-p-1)} \bar{\psi}_{\mathfrak{b}}^{(p)} \bar{\phi}_{\mathfrak{a}}^{(n)}} \Big)
\nl
+   c_1(m, n, m-n-p-1)  \varepsilon^{bc}  \Big( \state{\bar{\phi}_{\mathfrak{a}}^{(n)} \phi_b^{(p)} \psi_c^{(m-n-p-1)}} -  \state{\psi_c^{(m-n-p-1)} \phi_b^{(p)}  \bar{\phi}_{\mathfrak{a}}^{(n)}}\Big),
\nln
\hat{\gen{Q}}\state{\psi_a^{(m)}} \eq -\sum_{n+p < m} \bigg(c_2(m, n, p)\varepsilon^{\mathfrak{bc}}\Big(   \state{\psi_a^{(n)} \bar{\psi}_{\mathfrak{b}}^{(p)} \bar{\phi}_{\mathfrak{c}}^{(m-n-p-1)}}  +  \state{\bar{\phi}_{\mathfrak{c}}^{(m-n-p-1)} \bar{\psi}_{\mathfrak{b}}^{(p)} \psi_a^{(n)}} \Big)
\nl
+   c_2(m, n, m-n-p-1) \varepsilon^{bc}  \Big( \state{\psi_a^{(n)} \phi_b^{(p)} \psi_c^{(m-n-p-1)}} +  \state{\psi_c^{(m-n-p-1)} \phi_b^{(p)}  \psi_a^{(n)}}\Big) \bigg)
\nl
- \sum_{n+p \leq m} c_3(m, n, p) \varepsilon^{\mathfrak{bc}} \state{\bar{\phi}_{\mathfrak{b}}^{(n)}\phi_a^{(p)}\bar{\phi}_{\mathfrak{c}}^{(m-n-p)}}.
\>
With the normalization $g_3=1/2$, $c_1$ becomes
\[
c_1(m,n, p) =  \frac{ r_-(m)}{\sqrt{2}(m-n)r_-(n)r_+(p)r_-(m-n-p-1)}.
\]
The next coefficient function takes a very similar form, just switching some $r_+$ and $r_-$ and an overall minus sign, 
\[
c_2(m,n, p) =  -\frac{ r_+(m)}{\sqrt{2}(m-n)r_+(n)r_+(p)r_-(m-n-p-1)}.
\]
Finally, the last coefficient function takes a similar form, without the $\sqrt{2}(m-n)$ factor in the denominator,
\[
c_3(m,n, p) =  -\frac{ r_+(m)}{r_-(n)r_-(p)r_-(m-n-p)}.
\]
Initial (final) fermions lead to $r_+$ factors in the numerator (denominator) and initial (final)  scalars to $r_-$ factors in the numerator (denominator). 

Recall that the $\alg{su}(1|1)$ supercharges are nilpotent. One could check that $\hat{\gen{Q}}^2=0$ by working out all of its one-to-five site interactions in terms of the $c_i$. However, this is redundant; a few basic facts about the spin modules plus the vanishing commutators (up to gauge transformations) of $\hat{\gen{Q}}$ with the $\alg{osp}(4|2)$ generators already ensure that $\hat{\gen{Q}}^2=0$. We leave it as an exercise for the reader to work out this argument.

Manifest Hermiticity of the leading $\alg{osp}(4|2)$ generators implies that above we could  have considered $\hat{\gen{S}}$ instead of $\hat{\gen{Q}}$, and we would have then found a unique solution to $\hat{\gen{S}}$ (up to normalization). This was anticipated by the previously stated equality $\hat{\gen{S}} = (\hat{\gen{Q}})^\dagger$. Using this equality it is straightforward to work out the interactions of $\hat{\gen{S}}$ by switching initial and final states of (\ref{eq:qhatsolution}-\ref{eq:qhatsolutionconjugate}). For instance, the first line of (\ref{eq:qhatsolution}) implies
\<
\hat{\gen{S}}\state{\phi_a^{(m)}\psi_b^{(n)}\phi_c^{(p)}} \eq  \phantom{+}c_1(m+n+p+1, m, n)\varepsilon_{bc}\state{\phi_a^{(m+n+p+1)}}
\nl
+c_1(m+n+p+1, p, n)\varepsilon_{ab}\state{\phi_c^{(m+n+p+1)}}.
\>
The complete solution for  $\hat{\gen{S}}$ is given in the light-cone basis in Appendix \ref{sec:lightcone}.

From the expression for $\hat{\gen{Q}}$, we see that the (two-loop) $\alg{osp}(4|2)$ sector has an additional discrete symmetry under \emph{spin-chain} parity $\mathbf{p}$, which reverses the order of the spin chain sites with an extra signs for each crossing of fermions. After application of parity each site will have the opposite type of representation.	Note that $\mathbf{p}$ is distinct from space-time parity, and, unlike the case of  $\mathcal{N}=4$ SYM, this operation is distinct also from charge conjugation symmetry. While the leading order $\alg{osp}(4|2)$ generators are (trivially) parity even, $\hat{\gen{Q}}$ and $\hat{\gen{S}}$ are parity odd\footnote{Here we have chosen not to define $\mathbf{p}$ with a factor of $(-1)^L$ for chains with $2L$ sites, which would make all generators parity even.}.

\subsection{Solution for $\hat{\gen{Q}}$ in light-cone superspace basis}
In the light-cone basis the $r_{\pm}$ factors are absorbed into the normalization of the states in the expansion of $\state{x, \theta, \bar{\eta}}$, while the $(m-n)$ factors in the denominators can be accounted for with an integral as
\< \label{eq:lightconeqhat}
\hat{\gen{Q}} \state{x, \theta, \bar{\eta}} \eq \int_0^x \mathrm{d}y \,  \Big( \varepsilon_{ab} \partial^a_2 \partial^b_3  +  \varepsilon_{\mathfrak{ab}} \partial^{\mathfrak{a}}_2  \partial^{\mathfrak{b}}_3 \Big) \state{x, \theta, \bar{\eta}; y, \bar{\theta}_2, \eta_2;y, \theta_3, \bar{\eta}_3}   
 \nl
 + \int_0^x \mathrm{d}y \,  \Big( \varepsilon_{ab}  \partial^a_1 \partial^b_2   +  \varepsilon_{\mathfrak{ab}}  \partial_1^{\mathfrak{a}} \partial_2^{\mathfrak{b}}  \Big) \state{ y, \theta_1, \bar{\eta}_1; y, \bar{\theta}_2, \eta_2; x, \theta, \bar{\eta}} 
 \nl
  -\varepsilon_{\mathfrak{ab}} \bar{\eta}^{\mathfrak{a}} \partial_2^{\mathfrak{b}} \varepsilon_{cd}  \partial_1^c\partial_3^d    \state{ x, \theta_1, \bar{\eta}_1;x, \bar{\theta}_2, \eta_2; x, \theta_3, \bar{\eta}_3 } . 
 \>
Here the subscripts label on which site the partial derivatives act. For example, in the first term $\partial_2^a=\varepsilon^{ac} \partial/\partial \eta_2^c$.  The expression for acting on $\mathcal{V}_{\bar \phi}^{(4|2)}$ just follows from switching $\alg{su}(2)$ indices in epsilon tensors and in the derivatives ($\partial^a \leftrightarrow \partial^{\mathfrak{a}}$),  and switching all states with their conjugates ($\state{x, \theta, \bar{\eta}} \leftrightarrow \state{x, \bar{\theta}, \eta}$). Also, there is an extra overall minus sign for acting on $\mathcal{V}_{\bar \phi}^{(4|2)}$. As an example, we check one term. Expanding the $\theta$ component of the first term of the first line of (\ref{eq:lightconeqhat}) we find
\< \label{eq:checklightconeqhat}
\hat{\gen{Q}} \state{x, \theta, \bar{\eta}}_{\theta} \eq \int_0^x \mathrm{d}y \,   \varepsilon_{ab} \partial^a_2 \partial^b_3   \state{x, \theta, \bar{\eta}; y, \bar{\theta}_2, \eta_2;y, \theta_3, \bar{\eta}_3} _{\theta}
\nln
\eq \int_0^x \mathrm{d}y \, \sum_{m_1, m_2,m_3} \frac{\sqrt{(2m_1)!(2m_2+1)! (2m_3)!}}{m_1!m_2!m_3!} x^{m_1}y^{m_2+m_3} \theta^a \varepsilon^{bc} \state{\phi_a^{(m_1)}\psi_b^{(m_2)}\phi_c^{(m_3)}}
\nln
\eq \sum_{n+p < m}  \frac{\sqrt{(2n)!(2p+1)! (2(m-n-p-1))!}}{n!p!(m-n-p-1)!} \frac{x^m \theta^a }{m-n}  \varepsilon^{bc}  \state{\phi_a^{(n)}\psi_b^{(p)}\phi_c^{(m-n-p-1)}}
\nln
\eq \sum_m  \frac{\sqrt{(2m)!}x^m}{m!} \theta^a \sum_{n+p < m}  c_1(m, n, p)   \varepsilon^{bc} \state{\phi_a^{(n)}\psi_b^{(p)}\phi_c^{(m-n-p-1)}}.
\>
To reach the third line we did the integral, substituted for $m_3$ using $m=m_1+m_2+m_3+1$, and then replaced $m_1,m_2$ with $n, p$. The combinatoric factor simplifies to $\sqrt{(2m)!}c_1(m, n, p)/m!$, yielding the last line after reordering factors. On the other hand, acting with the first term of the first line of (\ref{eq:qhatsolution}) on the left side of (\ref{eq:checklightconeqhat}) clearly leads to the same result, as needed. One can the check remaining terms, involving also those with initial fermions, in a similar fashion.  

Because the light-cone basis is not manifestly Hermitian, $\hat{\gen{S}}$ takes a more involved form (requiring integration over two auxiliary variables, rather than just one), again see Appendix  \ref{sec:lightcone}.

\subsection{Hamiltonian, wrapping interactions, and twist-one spectrum}
The two loop dilatation generator for the $\alg{osp}(4|2)$ sector now follows from the anticommutator (\ref{eq:qhatshatacomm})
\[
 \acomm{\hat{\gen{Q}}}{\hat{\gen{S}}} = \delta \gen{D}_2. \label{eq:ham}
\]
Since $\hat{\gen{Q}}$ and $\hat{\gen{S}}$ are spin-chain parity odd, the Hamiltonian is parity even. We will not explicitly compute the expansion of the Hamiltonian in terms of interactions, since there is a simpler way to prove integrability. Still, we have used (\ref{eq:ham}), (\ref{eq:qhatsolution}-\ref{eq:qhatsolutionconjugate}), and \texttt{Mathematica} to check the spectrum for many (cyclic) spin chain states of relatively low dimension and length, finding complete agreement with the Bethe ansatz predictions of \cite{Minahan:2008hf}. This is empirical confirmation of the integrability that we will prove in the next section. 

It is important that there are physically equivalent expressions for the Hamiltonian which have different expansions in terms of local interactions. For periodic spin chains, including the cyclic spin chains of this work, there is always freedom to add chain derivatives to spin-chain generators including the Hamiltonian. Chain derivatives are nonzero interactions that vanish on periodic states. An example three-site chain derivative acts on a periodic spin chain of length $2L$ as
\[
\sum_{i=1}^{L} \Big(\gen{L}(2i-1)\gen{X}(2i,2i +1)  -\gen{X}(2i,2i +1)\gen{L}(2i+2)\Big),
\]
where again $\gen{L}$ is the length generator which just gives $1/2$ when acting on an individual site. $\gen{X}$ here can be an arbitrary (length-preserving) two-site generator . If $\gen{X}$ is fermionic, though, there would be extra signs. Of course, there is another chain derivative where $\gen{X}$ acts on sites $(2i-1, 2i)$ instead. Similarly, there are physically equivalent expressions that act on a (superficially)  different number of sites, due to interactions including spectator sites. For example a two-site generator $\gen{Z}$ can be written equivalently as a three-site generator,
\[
\sum_{I=1}^{2L} \gen{Z}(i, i+1) = 2 \sum_{i=1}^{2L} \Big(\gen{Z}(i, i+1) \gen{L}(i+2)\Big).
\]

Importantly (\ref{eq:qhatshatacomm}) implies that there is no wrapping contribution for $\mathcal{N}=6$ Chern-Simons  until four loops, as is also the case for $\mathcal{N}=4$ SYM \cite{Beisert:2005fw,Kotikov:2007cy}. Each $\alg{osp}(6|4)$ highest weight two-site state is in one-to-one correspondence with a descendant which is an  $\alg{osp}(4|2)$ sector highest-weight. Furthermore, in this sector ${\hat{\gen{S}}}$ annihilates two-site states and $\hat{\gen{Q}}$ combines non-BPS states with four-site states in long multiplets for $\lambda \neq 0$. Wrapping interactions for $\hat{\gen{Q}}$ first appears when there are 3-to-5 site interactions, i.e. at $\mathcal{O}(\lambda^3)$, which leads to a wrapping contribution to the four-loop dilatation generator. 

In fact, it is straightforward to compute the anomalous dimensions of  two-site states at two loops. These states have twist one, and according to (\ref{eq:twositeCartan}) there is one  highest-weight twist-one state for each nonnegative integer $\alg{osp}(4|2)$ spin $s$. A convenient representative of the $s$th multiplet is in the $\alg{sl}(2)$ sector and has Lorentz spin $(s+1/2)$,
\[
\state{\Psi_s}= \sum_{m=0}^s (-1)^m \sqrt{\binom{2 s +1}{2m}} \state{\phi_1^{(m)} \psi_1^{(s-m)}}.
\]
$\state{\Psi_s}$ is annihilated by $\gen{J}^{22}$, and is therefore an eigenstate of the dilatation generator since there are no other such states with the same quantum numbers.  Then the two-loop contribution to the anomalous dimension,  $\Delta_{s,2}$, simply equals the coefficient of $\state{\phi_1^{(s)}\psi_1^{(0)}}$ for $\mathcal{H}\state{\Psi_s}$ divided by $(-1)^s \sqrt{2s+1}$. Since ${\hat{\gen{S}}}$ annihilates two-site states, the Hamiltonian reduces to $ \hat{\gen{S}}\hat{\gen{Q}}$. 

We can organize the contributions of $\hat{\gen{S}}\hat{\gen{Q}}$  as follows. There are (diagonal) terms from the product acting only on the first site, or only on the second site. As explained in Appendix \ref{sec:lightcone}, the one-site part of the Hamiltonian is $2 S_1(2m)$ for  $\state{\phi^{(m)}}$ and $2 S_1(2m +1)$ for $\state{\psi^{(m+1)}}$. Also, there is a contribution from the fermion-to-three-boson interaction of $\hat{\gen{Q}}$  combined with the conjugate interaction of $\hat{\gen{S}}$,  
\[
 \state{\phi_1^{(0)} \psi_1^{(s)}} \rightarrow \varepsilon^{\mathfrak{bc}} \state{\phi_1^{(0)} \bar{\phi}_{\mathfrak{b}}^{(0)}\phi_1^{(s)}\bar{\phi}_{\mathfrak{c}}^{(0)}} \rightarrow    \state{\phi_1^{(s)} \psi_1^{(0)}}.
  \]
Note that the second arrow refer to $\hat{\gen{S}}$ acting on the last and first two sites,  $\hat{\gen{S}}(4,1,2)$.  Finally, for the generic terms, $\hat{\gen{Q}}$ inserts a $\state{\psi_1^{(0)}}$ (and two additional module elements) and $\hat{\gen{S}}$ replaces the other three sites with $\state{\phi_1^{(s)}}$, yielding a multiple of $\state{\phi_1^{(s)}\psi_1^{(0)}}$. 
Combining all of these contributions yields the $s$th two-loop anomalous dimension,
\< \label{eq:delta2sum}
\Delta_{s,2} \eq  \sum_{m=0}^{s} \frac{(-1)^m \sqrt{\binom{2s+1}{2 m}}}{{-1}^s \sqrt{2 s+1}} A(m),
\nln
A(m) \eq 2 \, \delta_{ms} \big(S_1(2s) + S_1(1)\big) - \frac{2 \, \delta_{m0}}{\sqrt{2s+1}}   + \sum_{n=0}^{m-1} 2 \, c_1(m,n, 0)c_1(s, n, s-m) 
\nl
+ \sum_{n=0}^{s-m-1} \Big(- 2 \,c_2(s-m, 0, n) c_1(s, s-m-n-1, n) 
\nl
+  8\, c_2(s-m, 0, n)c_1(s, m, n) + 2\, c_2(s-m, n, 0)c_1(s, m, n)  \Big). 
\>
It is simple to find the pattern numerically by evaluating the sum for low values of $s$, but in fact we have also done the sum analytically for arbitrary (nonnegative integer) $s$. For this it is significantly easier to use the light-cone superspace expressions for the supercharges, and we give more details about this in Appendix \ref{sec:lightcone}. We  find the spectrum in terms of the harmonic numbers and a generalized harmonic sum,
\[ \label{eq:twistonespectrum}
\Delta_s = 4 \lambda^2 \big(S_1(s) - S_{-1}(s)\big) + \mathcal{O}(\lambda^3),
\]
which is similar to the twist-two spectrum of $\mathcal{N}=4$ SYM, $8 \lambda_{\mathcal{N}=4} S_1(s)$. 

As we will see,  the R-matrix construction requires at least four sites. Still the Bethe ansatz correctly gives this twist-one spectrum because it naturally accounts for the action of $\hat{\gen{Q}}$ on two-site states described earlier. For zero-momentum solutions of the Bethe ansatz equations of \cite{Gromov:2008qe}\footnote{With grading $\eta=1$. For the opposite grading, one adds rather than removes a $u_3$ root at zero.}, $\hat{\gen{Q}}$ acts via $L \rightarrow (L+1)$ simultaneous with the removal of a single $u_3$ root at $0$, with all other roots unchanged. It is straightforward to check that this transformation does not change the energy or momentum (zero), and that it carries the same Cartan charges as $\hat{\gen{Q}}$ does. It follows that the spectrum for $L=1$ states is the same as the spectrum of $L=2$ states without a $u_3$ root at zero (and only $u_4$, $\bar{u}_4$ and $u_3$ roots excited). A similar phenomenon occurs in the $\alg{psu}(1,1|2)$ sector of $\mathcal{N}=4$ SYM \cite{Beisert:2005fw}.

In Appendix \ref{sec:bethe} we check analytically that the Bethe ansatz prediction gives precisely (\ref{eq:twistonespectrum}). This is already very strong evidence in favor of the leading-order Bethe equations of \cite{Minahan:2008hf}.  As found by \cite{Gromov:2008qe}, the large $s$ behavior of $\Delta_s$ gives a cusp anomalous dimension of $f(\lambda) = 4 \lambda^2$, though this disagrees by a factor of four with the value give in \cite{Aharony:2008ug}, based on \cite{Gaiotto:2007qi}. The author does not know the origin of this discrepancy, which was already noted in \cite{Gromov:2008qe}.

\section{Proof of $\alg{osp}(4|2)$ sector integrability \label{sec:yangian}}
In this section we prove that the two-loop dilatation generator for the $\alg{osp}(4|2)$ sector is integrable by constructing an $\alg{osp}(4|2)$ Yangian that commutes with the leading-order $\alg{su}(1|1)$ generators, and therefore with the two-loop dilatation generator. 
\subsection{Leading order Yangian}
A Yangian was used in a similar context for $\mathcal{N}=4$ SYM in \cite{Dolan:2003uh}. This is an infinite-dimensional symmetry algebra generated by the ordinary Lie algebra generators $\gen{J}^A$ and bilocal products of Lie algebra generators 
\[ \label{eq:definey}
\gen{Y}^A = f^{A}{}_{CB} \sum_{i<j} \gen{J}^B(i) \gen{J}^C(j),
\]
where $f^{A}{}_{CB}$ are the structure constants, with indices lowered (raised) using the (inverse) Cartan-Killing form. Note that the  $\gen{Y}^A $ are incompatible with periodic boundary conditions. As a result, generically the Yangian symmetry is only unbroken for infinite-length chains, as will be the case here.  The $ \gen{Y}^A$ manifestly transform in the adjoint of the Lie algebra. To consistently generate the entire algebra, the only additional requirement is that these $\gen{Y}^A$ satisfy Serre relations 
\[ \label{eq:serre}
3 \gcomm{\gen{Y}^{[A  }}{\gcomm{\gen{J}^B}{\gen{Y}^{C\}}}}  =  -  (-1)^{(EM)} f^{AK}{}_{D} f^B{}_{E}{}^{L} f^C{}_{F}{}^{M}  f_{KLM}\{ \gen{J}^D, \gen{J}^E, \gen{J}^F \},
 \]
where the curly brackets on the right side refer to the totally symmetric triple product, including a factor of $1/6$. Note that the indices on the left side are anti-symmetrized, with a factor of $1/6$. The signs and ordering of indices in the structure constants properly account for fermionic statistics for super Yangians \cite{Zwiebel:2006cb}. For $\alg{osp}(4|2)$, a possible basis is  $\{\gen{J}^A \} = \{\gen{Q}^{a\beta\mathfrak{c}},\, \gen{R}^{ab}, \tilde{\gen{R}}^{\mathfrak{ab}}, \gen{J}^{\alpha\beta} \}$. 

As reviewed in \cite{Dolan:2004ps}, it is sufficient to check that the Serre relations are satisfied for a one-site chain. This is because the Yangian, a Hopf algebra, has a coproduct which gives the Yangian's action on tensor products ((\ref{eq:definey}) actually follows from the coproduct). If the Yangian's Serre relations are satisfied on a single module, because of the coproduct they will be satisfied on chains of arbitrary length.  For a one-site chain, the left-side of the Serre relation (\ref{eq:serre}) vanishes, so we simply need to confirm that the right side vanishes. Programming the generators in $\texttt{Mathematica}$, we have confirmed that the Serre relations are satisfied when acting on any element of a single $\alg{osp}(4|2)$ module (of either type)\footnote{We have  obtained extra confirmation by also performed a number of checks of the Serre relation on two-site alternating or homogeneous chains. Of course, as stated above, the Serre relations are guaranteed to be satisfied because of the one-site result.}, as required.

\subsection{Vanishing commutator with $\hat{\gen{Q}}$ and $\hat{\gen{S}}$}
We will now explicitly show that the Yangian generators $\gen{Y}^{aa}$, which have the same $\alg{osp}(4|2)$ indices as $\gen{R}^{11}$ or $\gen{R}^{22}$, commute with  $\hat{\gen{Q}}$ on infinite-length chains. At the end of this section we will infer from this that the $\alg{osp}(4|2)$ Hamiltonian is integrable. 

First, consider the commutators between $\hat{\gen{Q}}$ and the $\alg{osp}(4|2)$ Lie algebra generators. Because $\gen{R}$ symmetry is manifest, the commutator of $\gen{R}$ with $\hat{\gen{Q}}$ vanishes locally, not just up to a gauge transformation. Also,  $\gen{Q}^{a2\mathfrak{b}}$ commutes with $\hat{\gen{Q}}$ locally as well (this commutator has classical dimension $0$, and a gauge transformation that inserts two sites has a minimum dimension $1$, twice the dimension of scalars without derivatives). On the other hand, the commutator with $\gen{Q}^{a1\mathfrak{b}}$ gives the gauge transformation (\ref{eq:qqhatgauge}) that acts as 
\<
\acomm{\gen{Q}^{a1\mathfrak{b}}}{\hat{\gen{Q}}} \eq \sum_i \grave{Z}^{a\mathfrak{b}}_i -\acute{Z}^{a\mathfrak{b}}_i ,
\nln
\grave{Z}^{a\mathfrak{b}}\state{x, \theta, \bar{\eta}} \eq \partial^a_1\bar{\partial}^{\mathfrak{b}}_2 \state{0, \theta_1, \bar{\eta}_1;0, \bar{\theta}_2, \eta_2;x, \theta, \bar{\eta}},
\nln
\acute{Z}^{a\mathfrak{b}}\state{x, \theta, \bar{\eta}} \eq \bar{\partial}^{\mathfrak{b}}_2\partial^a_3 \state{x, \theta, \bar{\eta}; 0, \bar{\theta}_2, \eta_2; 0, \theta_3, \bar{\eta}_3;},
\nln
\grave{Z}^{a\mathfrak{b}}\state{x, \bar{\theta}, \eta} \eq \bar{\partial}^{\mathfrak{b}}_1\partial^a_2 \state{0, \bar{\theta}_1, \eta_1;0, \theta_2, \bar{\eta}_2;x, \bar{\theta}, \eta},
\nln
\acute{Z}^{a\mathfrak{b}}\state{x, \bar{\theta}, \eta} \eq \partial^a_2 \bar{\partial}^{\mathfrak{b}}_3\state{x, \bar{\theta},\eta; 0, \theta_2, \bar{\eta}_2; 0, \bar{\theta}_3, \eta_3;}.
\>
In terms of components, the $\acute{Z}$ and $\grave{Z}$ insert two scalars without derivatives on adjacent sites.  Our convention for site indices is that $\hat{\gen{Q}}_i$, $\acute{Z}_i$ or $\grave{Z}_i$  or act on site $i$ and inserts sites $i+1$ and $i+2$ .  Therefore, the initial (and final) sites $1$ through $i-1$ are unaffected by $\hat{\gen{Q}}_i$, while for $j >i$, an initial site $j$ becomes site $j+2$, with these sites otherwise unchanged.

Before focusing on the $\gen{Y}^{aa}$, we consider general features of the commutator between $\hat{\gen{Q}}$ and bilocal generators. Consider the commutator involving one-site (bosonic) generators $\gen{J}^A$ and $\gen{J}^B$,
\[ \label{eq:commbilocalqhat}
\comm{\sum_{i < j} \gen{J}_i^A \gen{J}_j^B}{\hat{\gen{Q}}} = \sum_{i < j} \gen{J}_i^A \comm{\gen{J}^B}{\hat{\gen{Q}}}_j + \sum_{i < j}  \comm{\gen{J}^A}{\hat{\gen{Q}}}_i \gen{J}_j^B + \text{local}.
\]
Since $\hat{\gen{Q}}$ has one-to-three site interactions, the commutator with an individual $\gen{J}$ also gives a (spin-chain-local) one-to-three site generator, for which the subscript refers to the single site on which this new local generator acts. The terms of (\ref{eq:commbilocalqhat}) emerge as follows. The commutator vanishes when the $\gen{J}$ act on sites that $\hat{\gen{Q}}$ does not act on or insert. The terms where $\gen{J}^B$ acts on a site inserted or acted on by $\hat{\gen{Q}}$ but $\gen{J}^A$ acts completely to the left of $\hat{\gen{Q}}$ simplifies to the first term of the right side of (\ref{eq:commbilocalqhat}). The reflected terms, with $\gen{J}^A$ and $\gen{J}^B$ and right and left switched, yield the second term. Finally,  there are terms on which both $\gen{J}$ act on sites inserted by $\hat{\gen{Q}}$, which we call local because these combine into a homogeneous one-to-three site interaction
\[ \label{eq:local}
\sum_{i} \Big( \gen{J}^A_i (\gen{J}^B_{i+1} +\gen{J}^B_{i+2} ) +\gen{J}^A_{i+1}\gen{J}^B_{i+2} \Big) \gen{\hat{Q}}_i.
\]
Finally we turn to the $\gen{Y}^{aa}$. From (\ref{eq:definey}) we find
\<
4 \gen{Y}^{aa} \eq 
\sum_{i < j} 2 \varepsilon_{bc} \gen{R}^{bc}_i \gen{R}^{ca}_j -  \varepsilon_{\mathfrak{bc}} \gen{Q}^{a1 \mathfrak{b}}_i \gen{Q}^{a2 \mathfrak{c}}_j + \varepsilon_{\mathfrak{bc}} \gen{Q}^{a2 \mathfrak{b}}_i \gen{Q}^{a1 \mathfrak{c}}_j
\nln
\eq \gen{Y}^{aa}_{\gen{R}} + \gen{Y}^{aa}_{\gen{Q}^1} + \gen{Y}^{aa}_{\gen{Q}^2}. \label{eq:Ycc}
\>
where the factor of $4$ on the left side is for convenience. Using (\ref{eq:commbilocalqhat}) and (\ref{eq:Ycc}), the commutators described above, and taking into account statistics, we find
\<
4\comm{\gen{Y}^{aa}}{\hat{\gen{Q}}} \eq \sum_{i < j} \varepsilon_{\mathfrak{bc}} \acomm{\gen{Q}^{a1\mathfrak{b}}}{\hat{\gen{Q}}}_i \gen{Q}_j^{a2\mathfrak{c}} - \sum_{i < j} \varepsilon_{\mathfrak{bc}} \gen{Q}_i^{a2\mathfrak{c}} \acomm{\gen{Q}^{a1\mathfrak{b}}}{\hat{\gen{Q}}}_j  + \text{local}
\nln
\eq
\sum_{i < j} \varepsilon_{\mathfrak{bc}} (\grave{Z}_i^{a\mathfrak{b}} -\acute{Z}_i^{a\mathfrak{b}})  \gen{Q}_j^{a2\mathfrak{c}} - \sum_{i < j} \varepsilon_{\mathfrak{bc}} \gen{Q}_i^{a2\mathfrak{c}}(\grave{Z}_j^{a\mathfrak{b}} -\acute{Z}_j^{a\mathfrak{b}})  + \text{local}
\nln
\eq - \sum_{i} \varepsilon_{\mathfrak{bc}}  \acute{Z}_i^{a\mathfrak{b}}  \gen{Q}_{i+1}^{a2\mathfrak{c}} - \sum_i \varepsilon_{\mathfrak{bc}} \gen{Q}_i^{a2\mathfrak{c}}\grave{Z}_{i+1}^{a\mathfrak{b}}   + \text{local}. \label{eq:bilocalYqhatcomm}
\>
To reach the last line, we used the cancellations between $\acute{Z}$ and $\grave{Z}$ acting upon adjacent sites, similar to the cancellation explained after (\ref{eq:gauge}). In Appendix \ref{sec:localycomm} we evaluate the remaining local piece, which involves computing (\ref{eq:local}) with the $\gen{J}$ there replaced with the generators that appear in (\ref{eq:Ycc}).  The result leads to the precise cancellation on infinite-length chains,
\[ \comm{\gen{Y}^{aa}}{\hat{\gen{Q}}}=0.
\]

Since $\hat{\gen{Q}}$  commutes with all $\alg{osp}(4|2)$ Lie algebra generators and because the Yangian generators transform in the adjoint of $\alg{osp}(4|2)$, this is sufficient to imply that all generators commute with $\hat{\gen{Q}}$. For instance, any of the fermionic Yangian generator 
can be written as
\[
\gen{Y}^{a\beta\mathfrak{c}} = \pm \comm{\gen{Q}^{d\beta\mathfrak{c}}}{\gen{Y}^{aa}}, \quad d \neq a,
\]
which then necessarily commute with $\hat{\gen{Q}}$ since the $\alg{osp}(4|2)$ supercharges do (on infinite-length chains, as needed). Similarly, one can extend this to the remaining Yangian generators.  Hermiticity then implies that $\hat{\gen{S}}$ also commutes with the Yangian. Therefore, the two-loop $\alg{osp}(4|2)$ sector dilatation generator
\[
\delta\gen{D}_2 =  \acomm{\hat{\gen{Q}}}{\hat{\gen{S}}} 
\]
has an $\alg{osp}(4|2)$ Yangian symmetry and is integrable.

\section{R-matrix construction of the $\alg{osp}(4|2)$ sector Hamiltonian \label{sec:rmatrix}}
The $\alg{osp}(4|2)$ sector spin chain and its Yangian symmetry can be restricted consistently to the $\alg{sl}(2)$ sector. For such a $\alg{sl}(2)$ alternating spin chain we can use the known universal $\alg{sl}(2)$ R-matrix  \cite{Kulish:1981gi} to construct a transfer matrix as a function of two spectral parameters, $u$ and $\alpha$. We will see below that in our case $\alpha=0$. As is well-known, the expansion about $u=\infty$ gives the Yangian symmetry, while the expansion about $u=0$ gives the complete set of local conserved charges, $\mathcal{Q}_I$, that commute with the Yangian on infinite-length chains. Then the Hamiltonian for the  $\alg{sl}(2)$ sector of the ABJM spin chain must be a linear combination of the $\mathcal{Q}_I$, up to physically irrelevant chain derivatives. The restriction to nearest- and next-nearest-neighbor interactions identifies this Hamiltonian uniquely, up to coefficients that can be fixed by acting on a few states. 

Since the $\alg{sl}(2)$ sector Hamiltonian originates from an R-matrix construction, we can conclude that the $\alg{osp}(4|2)$ sector Hamiltonian also follows from an R-matrix construction. The argument is as follows. In  Appendix \ref{sec:maptosl2} we show that there is a unique lift from the $\alg{sl}(2)$ sector Hamiltonian to the $\alg{osp}(4|2)$ sector.  The (Lie algebra invariant) R-matrix only acts on two sites at a time. Also, recall that there is a one-to-one map between irreducible modules in the tensor product of two-sites in the $\alg{sl}(2)$ subsector and those in the $\alg{osp}(4|2)$ sector. Then the $\alg{osp}(4|2)$ sector Hamiltonian must take the form that would follow from an $\alg{osp}(4|2)$ R-matrix construction (assuming the existence of such R-matrices).  The Yangian construction of the previous section confirms the existence of R-matrices for the $\alg{osp}(4|2)$ spin-chain modules.

In this section we review the general R-matrix construction of the transfer matrix and conserved charges for an alternating spin chain. Based on the above argument, we then apply this construction to the $\alg{osp}(4|2)$ sector. We deduce the action of the $\alg{osp}(4|2)$ R-matrix on the spin-chain modules from the universal $\alg{sl}(2)$ R-matrix, obtaining another expression for the Hamiltonian. This new expression for the Hamiltonian will enable us to obtain the full $\alg{osp}(6|4)$ two-loop dilatation generator in the next section.

 \subsection{The transfer matrix and local conserved charges}
 The following discussion parallels the recent construction for the alternating $\alg{su}(4)$ spin chain \cite{Minahan:2008hf}, and the original general construction of \cite{deVega:1991rc}. 
 
 We start with a R-matrix, which satisfies the Yang-Baxter equation
\[
R_{12}(u-v)R_{13}(u)R_{23}(v)= R_{23}(v)R_{13}(u)R_{12}(u-v), \label{eq:ybe}
\]
where $R_{ij}$ is the R-matrix acting on sites $i$ and $j$. For alternating chains, it is sufficient for the R-matrix to satisfy the Yang-Baxter for each of the $2^3=8$ possible ways to assign one of the two representations to sites $1,2,3$. 
   
Now we consider an alternating chain with the two representations distinguished by the presence or absence of a bar, $1, \bar{2} \ldots (2L-1), \overline{2L}$. We build two monodromy matrices from the R-matrix,
 \[
 \mathcal{T}_a(u, \alpha) = \prod_{i=1}^L R_{a,2i-1}(u)R_{a, \bar{2i}}(u+\alpha), \quad  \mathcal{T}_{\bar{b}}(u, \beta) = \prod_{i=1}^L R_{\bar{b},2i-1}(u+\beta)R_{\bar{b}, \bar{2i}}(u). 
 \]
Since the R-matrices satisfy the Yang-Baxter equation,  these monodromy matrices also satisfy the Yang-Baxter equations 
\<
R_{ab}(u-v) \mathcal{T}_a(u) \mathcal{T}_b(v)\eq \mathcal{T}_b(v) \mathcal{T}_a(u)R_{ab}(u-v),
\nln
R_{\bar{a}\bar{b}}(u-v) \mathcal{T}_{\bar{a}}(u) \mathcal{T}_{\bar{b}}(v) \eq \mathcal{T}_{\bar{b}}(v) \mathcal{T}_{\bar{a}}(u)R_{\bar{a}\bar{b}}(u-v).
\>
Moreover, if $\beta=-\alpha$, which we will choose from now on, the mixed Yang-Baxter equation is also satisfied
\[
R_{a\bar{b}}(u+\alpha-v) \mathcal{T}_{a}(u, \alpha)\mathcal{T}_{\bar{b}}(v, -\alpha)= \mathcal{T}_{\bar{b}}(v, -\alpha) \mathcal{T}_{a}(u,\alpha)R_{a\bar{b}}(u+\alpha-v).
\]
Taking the trace, and using the invertibility of the R-matrix, we infer that the transfer matrices,
\[
T(u, \alpha) = \mathrm{Tr}_a \mathcal{T}_a(u, \alpha), \quad \bar{T}(u, -\alpha) = \mathrm{Tr}_{\bar{b}} \mathcal{T}_{\bar{b}}(u, -\alpha),
\]
satisfy
\[
\comm{T(u, \alpha)}{T(v, \alpha)}=0, \quad \comm{\bar{T}(u, -\alpha)}{\bar{T}(v, -\alpha)}=0, \quad \comm{T(u, \alpha)}{\bar{T}(v, -\alpha)}=0.
\]
In particular, this implies the existence of (up to) $2L$ commuting generators.  The expansion of $T(u, \alpha)$ about $u=0$ gives $L$ commuting generators, which also commute with the $L$ commuting generators coming from the expansion of $\bar{T}(u, -\alpha)$ about $u=0$. However, we will only consider the first two terms in the expansions. The leading terms, the transfer matrices evaluated at zero spectral parameter, yield the generators
 \<
 \mathcal{Q}_1 \eq \prod_{i=1}^{L-1}R_{2i+3, 2i+1}(0) \prod_{i=1}^L R_{2i-1,2i}(\alpha) ,
 \nln
  \bar{\mathcal{Q}}_1 \eq \prod_{i=1}^{L}R_{2i-1, 2i}(-\alpha) \prod_{i=1}^{L-1} R_{2i,2i+2}(0). 
 \>
Here we have stopped including bars to distinguish representations, which are of one type for odd-numbered sites, and the other for even-numbered sites. Also, these expressions require that, when acting on two identical representations, at $u=0$ the $R$-matrix is proportional to the permutation generator, which will be the case for the R-matrices we consider. Simplifying the product using $R^{-1}(\alpha) = R(-\alpha)$, another property of our R-matrices, we obtain the two-site shift generator 
\[
\mathcal{Q}_1\bar{\mathcal{Q}}_1 =  \prod_{i=1}^{L-1}R_{2i+3, 2i+1}(0)\prod_{i=1}^{L-1} R_{2i,2i+2}(0). 
 \]

Expanding to $\mathcal{O}(u)$, the next charges are defined through 
\<
T(u)\eq \mathcal{Q}_1  + u \mathcal{Q}_1 \mathcal{Q}_2 + \ldots, 
\nln
\bar{T}(u) \eq \bar{\mathcal{Q}}_1  + u \bar{\mathcal{Q}}_1 \bar{\mathcal{Q}}_2 + \ldots
\>
The charges have nearest-neighbor and next-nearest-neighbor contributions
\[
\mathcal{Q}_2 = (\mathcal{Q}_2)_{NN} +  (\mathcal{Q}_2)_{NNN}, \quad \bar{\mathcal{Q}}_2 = (\bar{\mathcal{Q}}_2)_{NN} +  (\bar{\mathcal{Q}}_2)_{NNN}.
\]
The nearest-neighbor contribution can be chosen symmetrically as 
\[
(\mathcal{Q}_2)_{NN} =  \sum_{i=1}^{2L} R_{i,i+1}(-\alpha)R'_{i,i+1}(\alpha), \quad (\bar{\mathcal{Q}}_2)_{NN} = \sum_{i=1}^{2L} R_{i,i+1}(\alpha)R'_{i,i+1}(-\alpha)
\]
provided the next nearest-neighbor contributions are  
\<
(\mathcal{Q}_2)_{NNN} \eq  \sum_{i=1}^L \bigg( R_{2i-1, 2i}(\alpha)R_{2i-1, 2i+1}'(0)R_{2i-1, 2i+1}(0)R_{2i-1, 2i}(-\alpha) 
\nl
+  R_{2i, 2i+1}(-\alpha)R_{2i-1, 2i+1}'(0)R_{2i-1, 2i+1}(0)R_{2i, 2i+1}(\alpha)\bigg),
\nln
(\bar{\mathcal{Q}}_2)_{NNN} \eq  \sum_{i=1}^L \bigg(R_{2i, 2i+1}(-\alpha)R_{2i, 2i+2}'(0)R_{2i, 2i+2}(0)R_{2i, 2i+1}(\alpha) 
\nl
+  R_{2i+1, 2i+2}(\alpha)R_{2i, 2i+2}'(0)R_{2i, 2i+2}(0)R_{2i+1, 2i+2}(-\alpha)\bigg).
\>
To compute the next-nearest-neighbor terms we inserted a factor of $R(0)R^{-1}(0)=1$, and to obtain symmetric expressions we used  the  vanishing commutator between the $\mathcal{Q}_i$.  There are additional possibilities that differ by chain derivatives. 

\subsection{The $\alg{osp}(4|2)$ case}
As explained above, the  relevant $\alg{osp}(4|2)$ R-matrices are determined by the universal R-matrix of $\alg{sl}(2)$ \cite{Kulish:1981gi}. This also occurred for the one-loop $\mathcal{N}=4$ SYM spin \cite{Beisert:2003yb}.  The result is a  sum over the irreducible representations of the tensor product of two sites, labeled by $\alg{osp}(4|2)$ spin $j$, weighted by a certain ratio of Gamma functions, 
\[ \label{eq:sl2universal}
R_{12}(u) = \sum_{j} (-1)^j f(cu) \frac{\Gamma(j+1+c u)\Gamma(1- c u)}{\Gamma(j+1-c u)\Gamma(1+ c u)}\mathcal{P}^{(j)}_{12}.
\]
Here $\mathcal{P}^{(j)}$ is the projector that acts as the identity on $\alg{osp}(4|2)$ states with spin $j$, and gives zero on all other states.  Using (\ref{eq:42conjugateproduct}) and (\ref{eq:42likeproduct}), for the case of sites $1,2$ in alternate representations, the sum is over all nonnegative $j$, while for identical representations, $j$ takes values $n-\half$ for all nonnegative $n$. For our purposes, the function of the spectral parameter $f$ and the constant $c$ can be replaced simply with freedom in the normalization of the local charges. We set $c$ to 1 and choose $f$ to  cancel any overall factors of $\pm i$ from the $(-1)^j$ factors. The Yang-Baxter equation (\ref{eq:ybe}) is still satisfied even if we choose different (constant) values of $f$ for different pairs of representations. As in the previous section, with these conventions  $R^{-1}(u)= R(-u)$, and $R(0)$ acts as the permutation operator on identical representations.

We are almost ready to simply insert the expression (\ref{eq:sl2universal}) for the R-matrices into the expressions for $\mathcal{Q}_2$  and $\bar{\mathcal{Q}}_2$ given in the last subsection, but there are four coefficients to fix. These are $\alpha$, the coefficients of $\mathcal{Q}_2$ and $\bar{\mathcal{Q}}_2$, and the coefficient of the identity operator, which we are also free to add without spoiling integrability. In principle one could compute four eigenvalues to fix these coefficients. However, since the Hamiltonian is even under spin-chain parity, $\alpha$ must be zero. Also, symmetry under charge conjugation implies that $\mathcal{Q}_2$ and $\bar{\mathcal{Q}}_2$ have equal coefficients. We have found the final two coefficients by comparison with eigenvalues of the (previous expression for the) Hamiltonian.  $\mathcal{Q}_2$ and $\bar{\mathcal{Q}}_2$  have coefficient $\lambda^2/2$ and the identity has  coefficient $2 \lambda^2 \log 2$,
\<
\delta\gen{D}_2 \eq  \half (\mathcal{Q}_2 + \bar{\mathcal{Q}}_2)_{|\alpha=0} + 4 L \log 2 
\nln
\eq  \sum_{i=1}^{2L} \Bigg( R_{i, i+1}^{(0)}R'_{i, i+1}(0)  + 2 \log 2 + 
\nl
 \half \Big(R_{i, i+1}^{(0)} R_{i, i+2}^{(0)}R_{i, i+2}'(0)R_{i, i+1}^{(0)} + R_{i+1, i+2}^{(0)} R_{i, i+2}^{(0)}R_{i, i+2}'(0)R_{i+1, i+2}^{(0)}\Big) \Bigg). 
\>
$\mathcal{Q}_2$ and $\bar{\mathcal{Q}}_2$ combine nicely here, and we used the more compact notation $R^{(0)}$ for the R-matrix evaluated at zero spectral parameter, $R(0)$. 
Next, evaluating (\ref{eq:sl2universal}) and its derivative at $u=0$, the dependence on $j$ reduces to factors of $(-1)^j$ and the harmonic numbers, $S_1(j)$ \footnote{It is also possible to absorb the identity component into the next-nearest neighbor terms  since a permutation squared equals the identity, as does the sum (with unit weight) over the projectors for next-nearest neighbors. The coefficient would then be $(-1)^{j_1+j_3}(\half S_1(j_2-1/2) +  \log 2)$, which is rational.}, 
\< \label{eq:osp42projectorham}
\delta \gen{D}_2 \eq    \sum_{i=1}^{2L}  \Bigg(2 \log 2 + \sum_{j=0}^\infty S_1(j)\mathcal{P}^{(j)}_{i, i+1}  
\nl
 + \sum_{j_1,j_2,j_3=0}^\infty (-1)^{j_1+j_3} \half S_1(j_2-\half)  \Big(\mathcal{P}^{(j_1)}_{i, i+1} \mathcal{P}^{(j_2-1/2)}_{i, i+2}\mathcal{P}^{(j_3)}_{i, i+1}  
+ \mathcal{P}^{(j_1)}_{i+1, i+2} \mathcal{P}^{(j_2-1/2)}_{i, i+2}\mathcal{P}^{(j_3)}_{i+1, i+2}   \Big) \!\Bigg)\!. 
\nl
\>
 Using the explicit form for the projectors given in Appendix \ref{sec:projectors} and \texttt{Mathematica}, we have checked that this spin-chain Hamiltonian for the two-loop ($\alg{osp}(4|2)$ sector) dilatation generator leads to the correct two-magnon S-matrix, and that its spectrum for low numbers of excitations and short states agrees with Bethe ansatz predictions and  the alternative expression for the Hamiltonian as the anticommutator of $\hat{\gen{Q}}$ and $\hat{\gen{S}}$.  Still, the anticommutator of $\hat{\gen{Q}}$ and $\hat{\gen{S}}$ gives a slightly more general form that applies to two-site states also, while this R-matrix expression requires chains of length four.

\section{The lift to the complete $\alg{osp}(6|4)$ chain \label{sec:lift}}
Here we derive the two-loop planar $\alg{osp}(6|4)$ dilatation generator using superconformal invariance. We then observe that this spin-chain Hamiltonian is integrable, assuming the existence of an $\alg{osp}(6|4)$ R-matrix for like or conjugate spin-chain modules. We argue that there is no reason to doubt this assumption.

\subsection{Unique lift from $\alg{osp}(4|2)$ to $\alg{osp}(6|4)$ \label{sec:uniquelift}}
By adding chain derivatives to replace one-site or two-site interactions with three-site interactions, one can write the $\alg{osp}(6|4)$ two-loop Hamiltonian completely in terms of a Hamiltonian density $\mathcal{H}_{i,i+1,i+2}$ of three-site to three-site interactions\footnote{The absence of one- or two-site interactions slightly simplifies our argument, but is not essential.},
\[
\mathcal{H}\state{X_1\ldots X_{2L}} = \sum_{i=1}^{2L} \mathcal{H}_{i,i+1,i+2} \state{X_1\ldots X_{2L}}.
\]
$\alg{osp}(6|4)$ invariance allows us to use the freedom to add chain derivatives so the Hamiltonian \emph{density} commutes with the leading-order $\alg{osp}(6|4)$ generators. Therefore the Hamiltonian is completely specified by the Hamiltonian density's action on all three-site (without cyclicity condition) highest-weight states.  Furthermore, the Hamiltonian density mixes highest-weight states only with other highest-weight states with the same $\alg{su}(4)$ Cartan charges, classical dimension, and Lorentz spin. For fixed values of these five Cartan charges there are finitely many linearly-independent three-site highest-weight states,  $\state{\Omega_I}$. In the sector with basis $\state{\Omega_I}$, the Hamiltonian density acts as
\[
\mathcal{H}_{123} \state{\Omega_I} = C^J_I \state{\Omega_J},
\]
for some coefficients $C^J_I$. The full set of such $C^J_I$ gives the Hamiltonian density and the Hamiltonian. If the Cartan charges of the$ \state{\Omega_I}$ satisfy a BPS condition, then these states are in an $\alg{osp}(4|2)$ sector\footnote{There are twelve isomorphic $\alg{osp}(4|2)$ sectors from different choices of the $1/12$ BPS condition.}, and we have already determined the action of the Hamiltonian (density). If not, as we show below, one can act with a combination of supercharges $\prod \gen{Q}$ (determined by the Cartan charges only) so that 
\[
\big(\prod \gen{Q}\big) \state{\Omega_I} = M^{I'}_I \state{\Omega'_{I'}}, \label{eq:maptoosp42}
\]
where the $\state{\Omega_{I'}}$ are contained within an $\alg{osp}(4|2)$ sector and the matrix $M$ is invertible. We have
\[
\mathcal{H}_{123} \state{\Omega'_{I'}} = C'^{J'}_{I'} \state{\Omega'_{J'}},
\]
for coefficients $C'^{J'}_{I'}$ determined by the known $\alg{osp}(4|2)$ sector Hamiltonian. Then, the needed coefficients of the full Hamiltonian are given by
\[
 C =  M C' M^{-1}, \quad \text{or} \quad C^J_I = M^{I'}_{I} C'^{J'}_{I'} ((M)^{-1})^{J}_{J'}.
 \]
Therefore, as claimed, there is a unique lift of the $\alg{osp}(4|2)$ sector Hamiltonian to $\alg{osp}(6|4)$. We will give this Hamiltonian below, after first proving the existence of the invertible map (\ref{eq:maptoosp42}). In Appendix \ref{sec:maptosl2} we go one step further and use the same type of argument to show that there is a unique lift from the  $\alg{sl}(2)$ sector to the $\alg{osp}(4|2)$ sector (and therefore to $\alg{osp}(6|4)$).

\subsection{Invertible map between $\alg{osp}(6|4)$ and $\alg{osp}(4|2)$ states \label{sec:invertiblemap}}
We will first show that for any basis $\state{\Omega_I}$ of three-site highest-weight states with identical Cartan charges  there exists a product of supercharges that maps the $\state{\Omega_I}$ to a $\alg{osp}(4|2)$ sector.  It will be straightforward afterward to show that this map is invertible. 

Again we choose the roots of $\alg{osp}(6|4)$ so that highest-weight states are annihilated by the following raising generators,
\[
 \gen{L}^2_1, \quad  \gen{K},  \quad \gen{R}^i_j \, \,  i >j,  \quad \gen{S}. \label{eq:rraise}
\]
Act on the $\state{\Omega_I}$ with $\gen{Q}_{12,1}$. If the Cartan charges of the $\state{\Omega_I}$ satisfy the BPS condition for $\gen{Q}_{12,1}$ and $\gen{S}^{12,1}$, this will vanish, the $\state{\Omega}_I$ are in an $\alg{osp}(4|2)$ sector, and the needed map is trivial. So we can assume $ \gen{Q}_{12,1} \state{\Omega_I}\neq 0$ .

Since the highest-weight states $\state{\Omega_I}$ are annihilated by the $\gen{R}$ of (\ref{eq:rraise}) it is a simple exercise to check that now the $\state{\Omega_I}$ must transform as $\mathbf{4}$ or $\mathbf{\bar{4}}$ under $\gen{R}$. This is because $\gen{Q}_{12,1}$ has nonvanishing action only  on lower $\gen{R}$ indices $3$ or $4$ or upper indices $1$ or $2$, and because the $\state{\Omega_I}$ have spin-chain length three.  
For simplicity, assume the $\state{\Omega_I}$ transform as $\mathbf{4}$. With appropriate interchange of indices this argument can be repeated for the $\mathbf{\bar{4}}$ case. 

Next consider
\[
\gen{Q}_{13,1} \gen{Q}_{12,1} \state{\Omega_I}.
\]
If these vanish then the $\gen{Q}_{12,1}\state{\Omega_I}$ satisfy the BPS condition for $\gen{Q}_{13,1}$ and $\gen{S}^{13,1}$, and therefore $\gen{Q}_{12,1}$ gives the required map to a $\alg{osp}(4|2)$ sector\footnote{Note that either all or none of the $\gen{Q}_{12,1}\state{\Omega_I}$ satisfy the BPS condition since they all have the same Cartan charges.}.  To see this use the last commutation relation of (\ref{eq:anticommutators}) and the annihilation of the $\state{\Omega_I}$ by the $\gen{R}$ raising generator $\gen{R}^3_2$. Similarly, if the
\[
\gen{Q}_{14,1}\gen{Q}_{13,1} \gen{Q}_{12,1} \state{\Omega_I}
\]
vanish, $\gen{Q}_{13,1} \gen{Q}_{12,1}$ gives the map to an $\alg{osp}(4|2)$ subsector. 

Finally, assume that $\gen{Q}_{14,1}\gen{Q}_{13,1} \gen{Q}_{12,1} \state{\Omega_I} \neq 0$. With respect to $\alg{su}(4)$ the scalars of the modules could transform  as  (fundamental, anti-fundamental, fundamental) or the conjugate. Since the argument would be essentially the same in either case, we assume the first possibility. For Lorentz spin $s$ and classical dimension $N + 3/2$ , the most general possibility is
\[
\gen{Q}_{14,1}\gen{Q}_{13,1} \gen{Q}_{12,1} \state{\Omega_I} = \sum_{n_{i,j}} a_I(n_{i, j}) \state{\phi_1^{(n_{1,1}, n_{2,1})} \psi_1^{(n_{1,2}, n_{2,2})}\phi_1^{(N + s -n_{1,1} - n_{1,2},N-s -n_{2,1} - n_{2,2})}} \label{eq:genformdescendant} 
\]
where the $n_{i,j}$ are nonnegative integers such that all superscript arguments are also nonnegative integers consistent with spin statistics, and $a_I(n_{i,j})$ are some coefficients. Now $\gen{S}^{12,2}$ must still annihilate (\ref{eq:genformdescendant}), since it anticommutes with $\gen{Q}_{13, 1}$ and  $\gen{Q}_{14, 1}$  and gives $\gen{R}^2_1$ when anticommuted with $\gen{Q}_{12, 1}$. $\gen{S}^{12,2}$ acts on single-sites with lower $1$ indices as
\<
\gen{S}^{12,2} \state{\phi_1^{(n_1, n_2)}} \eq  b(n_2) \state{(\bar{\psi}^2)^{(n_1, n_2-1)}}, \quad  b(n_2) = 0 \Leftrightarrow n_2=0.
\nln
\gen{S}^{12,2} \state{\psi_1^{(n_1, n_2)}} \eq  c(n_2) \state{(\bar{\phi}^2)^{(n_1, n_2-1)}}, \quad  c(n_2) = 0 \Leftrightarrow n_2=0,
\>
where all that matters here are the quantum number of the states and whether the coefficients $b$ and $c$ are nonzero. From this it follows that  $\gen{S}^{12,2}$  annihilates (\ref{eq:genformdescendant}) only if all of the second Lorentz index excitation numbers $n_{2,1}$, $n_{2,2}$ and $N-s-n_{2,1}-n_{2,2}$ are zero. But then all of the states of (\ref{eq:genformdescendant}) are clearly in an $\alg{osp}(4|2)$ ($\alg{sl}(2)$) sector, and in this case $\gen{Q}_{14,1}\gen{Q}_{13,1} \gen{Q}_{12,1}$ gives the required map.

We have shown that applying a product of supercharges (zero, one, two, or three depending on the Cartan charges) gives a map to an $\alg{osp}(4|2)$ sector. This is as abbreviated in (\ref{eq:maptoosp42}). To show that this map is invertible, we simply need to show the linear independence of the 
\[
 \big(\prod \gen{Q}\big)\state{\Omega_I},
 \]
where as usual we are focusing on a linearly-independent basis given by three-site highest-weight states $\state{\Omega_I}$ with identical Cartan charges.  If there were some linear combination of the $\state{\Omega_I}$ that were annihilated by the relevant $ \prod \gen{Q}$,  the above construction implies this linear combination would satisfy another BPS condition. This is a contradiction because the BPS conditions depend only on the Cartan charges (and because the $\state{\Omega_I}$ are assumed linearly independent).  Therefore, the maps of (\ref{eq:maptoosp42}) to the $\alg{osp}(4|2)$ sector are invertible, completing the proof.

\subsection{The two-loop $\alg{osp}(6|4)$ spin-chain Hamiltonian and integrability}
Recall that there is a one-to-one mapping between highest-weight  two-site states in the $\alg{osp}(4|2)$ subsector and the full theory, and that these highest-weight states  have the same  value of $\alg{osp}(4|2)$ or $\alg{osp}(6|4)$ spin. Therefore, the unique lift is given by replacing projections onto $\alg{osp}(4|2)$ spin in (\ref{eq:osp42projectorham}) with projections onto the corresponding $\alg{osp}(6|4)$ spin.    Now the complete planar Hamiltonian is given by the same formal expression as for the $\alg{osp}(4|2)$ sector,
\<\label{eq:completeham}
\delta \gen{D}_2 \eq    \sum_{i=1}^{2L}  \Bigg( 2 \log 2 + \sum_{j=0}^\infty S_1(j)\mathcal{P}^{(j)}_{i, i+1}    
\nl
 + \sum_{j_1,j_2,j_3=0}^\infty (-1)^{j_1+j_3} \half S_1(j_2-\half)  \Big(\mathcal{P}^{(j_1)}_{i, i+1} \mathcal{P}^{(j_2-1/2)}_{i, i+2}\mathcal{P}^{(j_3)}_{i, i+1}  
+ \mathcal{P}^{(j_1)}_{i+1, i+2} \mathcal{P}^{(j_2-1/2)}_{i, i+2}\mathcal{P}^{(j_3)}_{i+1, i+2}   \Big) \!\Bigg)\!. 
\nl
\>
Of course, the differences from the $\alg{osp}(4|2)$ sector Hamiltonian are that the projectors act on the full $\alg{osp}(6|4)$ modules, and the $j_i$ correspond to $\alg{osp}(6|4)$ spin. Again, there is a sum over the spin-chain sites labeled by $i$, with $\mathcal{P}^{(j)}_{i,k}$ acting on sites $i$ and $k$. It would be nice to have the expressions for these projectors in components, extending those given for the $\alg{osp}(4|2)$ sector in Appendix \ref{sec:projectors}.

Assuming the existence of an $\alg{osp}(6|4)$ R-matrix acting on like or conjugate pairs of the two types of $\alg{osp}(6|4)$ modules, a parallel derivation to the one given for $\alg{osp}(4|2)$ in Section \ref{sec:rmatrix} would apply, and would lead to the complete Hamiltonian given in (\ref{eq:completeham}). Therefore, up to this assumption, we have shown that planar $\mathcal{N}=6$ superconformal Chern-Simons theory is integrable at two-loops. A similar assumption was used in the $\mathcal{N}=4$ SYM case \cite{Beisert:2003yb}.
 
As noted earlier, the existence of Yangian symmetry for the $\alg{osp}(4|2)$ sector implies the existence of the corresponding $\alg{osp}(4|2)$ R-matrices for the modules appearing in the spin chain. This is indication that there is no problem constructing R-matrices for $\alg{osp}$ algebras. It seems that it would be sufficient to confirm that the Serre relations (\ref{eq:serre}) are satisfied for the complete $\alg{osp}(6|4)$ modules, since that would imply the existence of a $\alg{osp}(6|4)$ Yangian and the corresponding R-matrices. However, such a check is beyond the scope of this work.   
 
In fact, it reasonable to assume even a \emph{universal} $\alg{osp}(6|4)$ R-matrix\footnote{I thank E. Ragoucy for helpful related discussions.}, which would give the R-matrix for arbitrary $\alg{osp}(6|4)$ representations. There is a general construction of universal R-matrices for Yangians of bosonic simple Lie algebras \cite{Khoroshkin:1994uk} later extended to  $\alg{sl}(m|n), m\neq n$ \cite{Stukopin:2005st}. The latter construction was further modified for a recent derivation of the leading-order spin-chain S-matrices of $\mathcal{N}=4$ SYM and ABJM \cite{Spill:2008yr}. It should be possible to apply a similar construction to $\alg{osp}$ algebras as well\footnote{A degenerate Cartan matrix is an obstruction to constructing a R-matrix, but that is not a problem for $\alg{osp}(6|4)$ since its Cartan matrix is invertible.}. 

Finally, we emphasize  that the invariance of the Hamiltonian density with respect to $\alg{osp}(6|4)$ only applies at leading order.  At the next order in $\lambda$, the Hamiltonian will only commute exactly with at most a proper subset of the $\alg{osp}(6|4)$ supercharges, while the commutators with the other supercharges will vanish only  when applied to cyclic alternating chains. This can be see already in the $\alg{osp}(4|2)$ sector. While the expression (\ref{eq:osp42projectorham}) for the Hamiltonian commutes with $\mathcal{O}(\lambda^0)$ $\alg{osp}(4|2)$ generators manifestly, it only commutes with the $\mathcal{O}(\lambda^1)$ supercharges $\hat{\gen{Q}}$ and $\hat{\gen{S}}$ up to gauge transformations. A similar phenomenon occurs in the $\alg{psu}(1,1|2)$ sector of $\mathcal{N}=4$ SYM \cite{Beisert:2007sk}. It can be traced back to the algebra of supersymmetry variations only closing up to gauge transformations, which appear for the spin-chain as (\ref{eq:qqhatgauge}). Importantly, these gauge transformations appear at subleading order and do not obstruct the leading-order $\alg{osp}(6|4)$ invariance used to lift the Hamiltonian to the full theory.

\section{Conclusions \label{sec:conc}} 

We have shown that the two-loop planar dilatation generator of $\mathcal{N}=6$ superconformal Chern-Simons theory is fixed by superconformal invariance up to overall normalization, and can be written compactly as in (\ref{eq:completeham}). Through a Yangian construction we have proved integrability for the $\alg{osp}(4|2)$ sector, and for the full model assuming the existence of an $\alg{osp}(6|4)$ R-matrix. This confirms the conjectured two-loop $\alg{osp}(6|4)$ Bethe equations of Minahan and Zarembo. We also analytically computed the twist-one spectrum of the model, both from the Bethe equations and from the dilatation generator, finding $\Delta_s= 4 \lambda^2(S_1(s)-S_{-1}(s)) + \mathcal{O}(\lambda^3)$.
 
It seems unreasonable to doubt complete two-loop integrability. Still, an explicit proof would be better. Constructing the  $\alg{osp}(6|4)$ Yangian that commutes with the Hamiltonian (\ref{eq:completeham}) may  be the simplest approach. 

Further algebraic constructions of spin-chain generators for the ABJM gauge theory are possible. It would be good to calculate the complete $\mathcal{O}(\lambda)$ generators, extending the calculation here of two $\mathcal{O}(\lambda)$  supercharges acting within the $\alg{osp}(4|2)$ sector. It would also be wonderful to obtain higher-loop corrections. For $\mathcal{N}=4$ SYM there is evidence of recursive structure to these corrections, at least in sectors of the theory \cite{Zwiebel:2008gr, Bargheer:2008jt}. \cite{Bargheer:2008jt} also argues that such recursive structure appears in a compact sector for the ABJM spin chain (assuming higher-loop integrability). Perhaps this recursive structure extends to the $\alg{osp}(4|2)$ sector, which could make an algebraic computation of the four-loop dilatation generator tractable. This would still require a direct field theory calculation of $h(\lambda)$, which appears in the one-magnon dispersion relation.

We found it useful to work with a representation of the spin module in terms of continuous variables, a light-cone superspace basis. We expect this and similar representations to be helpful for gaining new insights about the $\mathcal{N}=4$ SYM and the ABJM spin chains. 

It is straightforward to lift the $\alg{osp}(4|2)$ dilatation generator to the complete  $\alg{osp}(4|2)$ two-loop dilatation generator, including nonplanar corrections. $\hat{\gen{Q}}$ ($\hat{\gen{S}}$) has a unique nonplanar lift since it acts on one initial (final) module. The nonplanar two-loop dilatation generator than follows from the anticommutator (\ref{eq:qhatshatacomm}), and this should match the $\alg{su}(2) \times \alg{su}(2)$ subsector result of \cite{Kristjansen:2008ib}.  There is a very similar observation for the $\alg{psu}(1,1|2)$ sector of $\mathcal{N}=4$ SYM \cite{Zwiebel:2005er}. Also for $\mathcal{N}=4$ SYM, an analysis of  the gauge group structure of the Feynman diagrams that contribute to one-loop anomalous dimensions led to  a unique lift from the planar limit to the full nonplanar theory  \cite{Beisert:2003jj}. The product gauge group and bifundamental fields of ABJM probably make the corresponding two-loop lift here more difficult. Still, a lift from the nonplanar $\alg{osp}(4|2)$ sector to the full theory should be possible, especially given the one-to-one maps of Section \ref{sec:lift} between highest-weight three-site states of the $\alg{osp}(4|2)$ and $\alg{osp}(6|4)$ sectors.  

The $\alg{osp}(4|2)$ sector dilatation generator, including nonplanar corrections, should be useful for seeking the gauge duals of 1/12 BPS black holes in $AdS_4$. For recent related work and comments see \cite{Bhattacharya:2008bja}. It would also be interesting to calculate (nonplanar) anomalous dimensions of near ($1/12$) BPS states, as done previously \cite{Berkooz:2008gc} for near ($1/16$) BPS operators in $\mathcal{N}=4$ SYM.
 
Finally, this work's confirmation of weak-coupling integrability further motivates study of the many topics related to the integrability of the ABJM gauge theory and its string theory dual. As for $AdS_5/CFT_4$, these topics range far beyond the weak-coupling spin chain.

 \subsection*{Acknowledgements}
I am very grateful to Niklas Beisert, Andrei Belitsky, Diego Hofman, and Eric Ragoucy for discussions. I also thank Georgios Papathanasiou and Marcus Spradlin for pointing out important typos in Appendix D of the first version of this work.
 
\appendix
\section{ Additional light-cone superspace computations \label{sec:lightcone}}
Working in light-cone superspace given by (\ref{eq:definelightcone}) often simplifies calculations. We gave the actions of $\hat{\gen{Q}}$ in light-cone superspace in (\ref{eq:lightconeqhat}), and in this appendix we give the corresponding expression for $\hat{\gen{S}}$. We then use this expression to compute the one-site interactions of $\mathcal{H}$ and and the two-loop twist-one spectrum.  
\subsection{$\hat{\gen{S}}$ in light-cone superspace}
We abbreviate
\[
\state{x_1, \theta_1, \bar{\eta}_1;x_2, \bar{\theta}_2, \eta_2 ;
x_3, \theta_3, \bar{\eta}_3} = \state{1;\bar{2},3}.
\]
Then, $\hat{\gen{S}}$ acts as 
\< \label{eq:lightconeshat}
\hat{\gen{S}} \state{1;\bar{2},3} \eq  -\frac{1}{ 2 \pi} \int_0^1 \frac{\mathrm{d}t_1}{\sqrt{t_1}\sqrt{1-t_1}}  \int_0^1 \frac{\mathrm{d}t_2}{t_2\sqrt{1-t_2}} \times
\nl
\bigg(\Big( \varepsilon_{bc} \eta_2^b \theta_3^c \partial_{x_2} + \varepsilon_{\mathfrak{bc}}\bar{\theta}_2^{\mathfrak{b}} \bar{\eta}_3^{\mathfrak{c}}  \partial_{x_3} \Big)  \state{(1-t_2) x_1 + t_1 t_2 x_2 + (1-t_1) t_2x_3, \theta_1, (1-t_2) \bar{\eta}_1}
\notag \\
&+& \Big( \phantom{\Big(}\varepsilon_{bc} \theta_1^b \eta_2^c  \partial_{x_2} + \varepsilon_{\mathfrak{bc}}\bar{\eta}_1^{\mathfrak{b}}  \bar{\theta}_2^{\mathfrak{c}}  \partial_{x_1} \Big)  \state{ (1-t_1)t_2 x_1 +   t_1 t_2 x_2 + (1-t_2) x_3, \theta_3, (1-t_2) \bar{\eta}_3}
\notag \\
&+& \bar{\theta}^{\mathfrak{a}}_2 \bar{\partial}_{\mathfrak{a}} \varepsilon_{bc} \theta_1^b \theta_3^c \state{ t_1 t_2 x_1 +   (1-t_2) x_2 + (1-t_1) t_2 x_3,\theta,  t_2  \bar{\eta}} \bigg).
\nl
\>
To check this, first expand both sides according to (\ref{eq:definelightcone}). Using the Beta integral one can then show that this is equivalent to the action of $\hat{\gen{S}}$ given previously (the Hermitian conjugate of the $\hat{\gen{Q}}$ action (\ref{eq:qhatsolution})). Since the light-cone superspace basis is not manifestly Hermitian, here $\hat{\gen{S}}$ takes a more involved form than $\hat{\gen{Q}}$ does. To obtain the action on the conjugate state, $\state{\bar{1};2\;\bar{3}}$, remove the overall minus sign and  replace all $\theta,\eta$ with $\bar{\theta},\bar{\eta}$ and vice-versa (interchanging all Latin and Gothic indices).

\subsection{One-site interactions of $\mathcal{H}$}
Because physical spin-chain states are cyclic and have at least two sites, in general one can not uniquely classify interactions as  one-site interactions rather than two-site (or three-site) interactions. However,  we can isolate the contribution to $\mathcal{H}$  from $\hat{\gen{Q}}$ replacing one site with three new sites followed by $\hat{\gen{S}}$ replacing the same three sites with a single site. This is what we mean here by $\mathcal{H}_{\text{one site}}$.  Using the light-cone superspace expressions  (\ref{eq:lightconeqhat}) and (\ref{eq:lightconeshat}), we find the following expression for $ \mathcal{H}_{\text{one site}}\state{x, \theta, \bar{\eta}}$,
\begin{gather}
 \frac{1}{ 2 \pi} \int_0^1 \frac{\mathrm{d} t_1}{\sqrt{t_1}\sqrt{1-t_1}}  \int_0^1 \frac{\mathrm{d}t_2}{t_2\sqrt{1-t_2}} \int_0^x \mathrm{d}\!y \,  \times \notag \\
 \bigg( 2 (\partial_{y_2} + \partial_{y_3}) \state{(1-t_2)x + t_1t_2 y_2 + (1-t_1)t_2 y_3, \theta, (1-t_2) \bar{\eta}}_{|^{y_2,y_3=y}}
 \notag \\
+ 2 \Big(\bar{\eta}^{\mathfrak{c}}\bar{\partial}_{\mathfrak{c}} \partial_x  -\theta^c \partial_{c}  \partial_{y_2} \Big)\state{(1-t_1)t_2 x + t_1t_2y_2 + (1-t_2)y,\theta, (1-t_2) \bar{\eta}}_{|^{y_2=y}}
 \notag \\
 +  2 (\partial_{y_1} + \partial_{y_2}) \state{(1-t_1)t_2 y_1 + t_1t_2 y_2 + (1-t_2) x, \theta, (1-t_2) \bar{\eta}}_{|^{y_1,y_2=y}} \bigg)
 \notag \\ 
 +  \frac{1}{2 \pi} \int_0^1 \frac{\mathrm{d}t_1}{\sqrt{t_1}\sqrt{1-t_1}}  \int_0^1 \frac{\mathrm{d}t_2}{t_2\sqrt{1-t_2}} 2 \bar{\eta}^{\mathfrak{a}} \bar{\partial}_{\mathfrak{a}} \state{x, \theta, t_2\bar{\eta}}.
  \end{gather}
Simplifying and doing integrals of derivatives and Beta integrals yields
\<
  \mathcal{H}_{1} \eq \int_0^1 \frac{\mathrm{d}t_2}{t_2\sqrt{1-t_2}}
 \Big( 2  \state{x, \theta, (1-t_2) \bar{\eta}} -2  \state{(1-t_2)x , \theta, (1-t_2) \bar{\eta}}\Big)
 \nl
 + \frac{1}{  \pi} \int_0^1 \frac{\mathrm{d}t_1}{\sqrt{t_1}\sqrt{1-t_1}}  \int_0^1 \frac{\mathrm{d}t_2}{t_2\sqrt{1-t_2}}  \times 
 \nl
  \frac{ t_2}{1- t_2 + t_1 t_2} \Big(  \bar{\eta}^{\mathfrak{a}} \bar{\partial}_{\mathfrak{a}} -t_1 \Big) \Big(\state{x, \theta, (1-t_2) \bar{\eta}} -\state{(1-t_1)t_2 x, \theta , (1-t_2)\bar{\eta}} \Big) 
 \nl
  +  2 \bar{\eta}^{\mathfrak{a}} \bar{\partial}_{\mathfrak{a}} \state{x, \theta, \bar{\eta}}.
  \>
We now expand the light-cone basis in components, with the $x^m \theta$ terms giving the action on $\state{\phi^{(m)}}$ and the $x^m \bar{\eta}$ terms giving the action on $\state{\bar{\psi}^{(m)}}$. The integrand of the middle two lines then simplifies to a finite sum of products of powers of the $t_i$ and $(1-t_i)$, so that the integrals again reduce to Beta integrals. Doing these integrals and the sum, and combining with contributions from the first and last line finally yields the result
\[
\mathcal{H}_{\text{one site}}\state{\phi_a^{(m)}} = 2 S_1(2m) \state{\phi_a^{(m)}}, \quad \mathcal{H}_{\text{one site}}\state{\bar{\psi}_{\mathfrak{a}}^{(m)}} = 2 S_1(2m+1) \state{\bar{\psi}_{\mathfrak{a}}^{(m)}}.
\]
Of course, the same coefficients appear for the conjugate scalars and fermions. We also used the identity: 
\[
S_1(m-\half) + S_1(m) + 2 \log 2 = 2 S_1(2m),
\]
where the definition of the harmonic numbers that applies for nonintegers is the difference involving the digamma function 
\[
S(x) = \psi(x+1)-\psi(1).
\]

\subsection{ Two-loop twist-one spectrum}
Working in light-cone superspace using (\ref{eq:lightconeqhat}) and (\ref{eq:lightconeshat}), we find the parallel expression to (\ref{eq:delta2sum}),
\< \label{eq:delta2integral}
\Delta_{s,2} \eq \sum_{m=0}^{s} \frac{(-1)^m \tilde{A}(m)}{(-1)^s \sqrt{2 s+1}},
\nln
\tilde{A}(m) \eq \delta_{ms} \sqrt{2 s+1} (2 \, S(2s) + 2 \, S(1)) - \frac{2 \,  \delta_{m0}}{\sqrt{2s+1}} 
\nl
+ \int \frac{\mathrm{d}t_1 \mathrm{d}t_2}{\pi} \, \mathrm{d}y\, \frac{t_1^{s-m+\half}t_2^{s-m}}{\sqrt{1-t_1}\sqrt{1-t_2}} \frac{m}{\sqrt{2s+1}} \binom{2 s+1}{2 m}\big((1-t_1)t_2y +(1-t_2)\big)^{m-1} 
\nl
+ \int \frac{\mathrm{d}t_1 \mathrm{d}t_2}{\pi} \, \mathrm{d}y\, \frac{t_1^{m-\half}t_2^{m}\sqrt{1-t_1}}{\sqrt{1-t_2}}  \frac{s-m}{\sqrt{2s+1}} \binom{2 s+1}{2 m} (1-t_1t_2)^{s-m-1}y^{s-m-1} 
\nl
-  \int \frac{\mathrm{d}t_1 \mathrm{d}t_2}{\pi} \, \mathrm{d}y\, \frac{\sqrt{t_1}t_2^{s-m-1}(1-t_2)^{m-\half}}{\sqrt{1-t_1}} \frac{s-m}{\sqrt{2s+1}} \binom{2 s+1}{2 m} \big(t_1+(1-t_1)y\big)^{s-m-1}
\nl
- 4 \int \frac{\mathrm{d}t_1 \mathrm{d}t_2}{\pi} \, \mathrm{d}y\, \frac{\sqrt{t_1}t_2^{s-m-1}(1-t_2)^{m-\half}}{\sqrt{1-t_1}}  \frac{s-m}{\sqrt{2s+1}} \binom{2 s+1}{2 m} y^{s-m-1}.
\>
The ranges of integration variables are all zero to one. The first integral should only be included in the sum for $m>0$, and the last three only for $m < s$. To obtain (\ref{eq:delta2integral}) we expanded the light-cone basis in components and absorbed (most of) the wavefunction and light-cone basis normalizations into the integrands. Next, the $y$ integrals are elementary, and, using binomial expansions, all of the $t_i$ integrals can be done using the Beta integral. The sums can also be done\footnote{We have used \text{Mathematica} to evaluate these sums, symbolically as a function of $s$.}, provided we use the identity 
\[
 \frac{\sqrt{\pi}\Gamma(s+\half)}{2}\sum_{m=1}^s\frac{(-1)^{m}}{m\Gamma(m+3/2)\Gamma(s-m+\half)} = 2 S_{-1}(s) - S_1(s)  +  2  -\frac{(-1)^s}{s+\half}.
\]
An equivalent version for integer $s$ that is also valid for more general $s$ is
\[
\frac{{}_3F_2(1,\sfrac{3}{2},s+1; s+2, s+\sfrac{5}{2};1)}{(s+1)(2s+1)(2s+3)}=S_1(\frac{s-1}{2}) - S_1(\frac{s}{2}) + \frac{1}{s+\half}.
\]
We have proved this identity through a somewhat involved calculation. Key steps include considering the difference of the identity at $s=s'$ and at $s=s'+1$, using the series expansion of hypergeometric functions, reintroducing Beta integrals (as well as another elementary integral), and switching orders of summation and integration. It would be nice to find an elegant proof of this identity, or better, a more elegant way to evaluate  $\Delta_{s,2}$ directly from the light-cone superspace expressions for the supercharges.  In any case, the final result is as given in (\ref{eq:twistonespectrum}), 
\[
\Delta_{s,2} = 4 \Big(S_1(s)-S_{-1}(s)\Big).
\]

\section{Bethe ansatz solution for two-loop twist-one spectrum \label{sec:bethe}}
Here we work with the $\eta=-1$ (leading-order) Bethe equations of \cite{Gromov:2008qe}. The $\alg{sl}(2)$ sector highest-weight state with Lorentz spin $(s+1/2)$ has $s$ pairs of roots $u_{4, k}=\bar{u}_{4, k}$. Labeling both types of roots $u_k$, the Bethe equations reduce to
\[
\frac{u_k+i/2}{u_k-i/2}=\prod_{j}\frac{u_k-u_j-i}{u_k-u_j+i}.
\]
As done for the parallel calculation in $\mathcal{N}=4$ SYM \cite{Eden:2006rx} based on \cite{Korchemsky:1995be}, we introduce the Baxter polynomial
\[
Q_s(u) = C_s\prod_k(u-u_k)
\]
 that satisfies
\[
T_s(u)Q_s(u) = (u+i/2) Q_s(u+i)-(u-i/2)Q_s(u-i)
\]
for some auxiliary polynomial $T_s$.  $C_s$ is a $u$-independent normalization factor.  Matching powers of $u$ on each side requires $T_s$ to be independent of $u$, and for this equation to have a solution for all $s\geq0$  $T_s=(2 s+1)i$.
The solution to a difference equation of this form is a Meixner polynomial (closely related to Jacobi and Krawtchouk  polynomials) \cite{Koekoek:1996xx}, 
\[
Q_s(u)={}_2F_1(-s, iu + \half; 1; 2).
\]
The two-loop contribution to the anomalous dimension is given by 
\[
\sum_k \frac{2 \lambda^2}{u_k^2 + \quarter},
\]
(the factor of $2$ is for the $u_4$ and $\bar{u}_4$ roots), which is proportional to the ratio of the coefficient of $u$ to the constant term in $Q_s(u+i/2)$.  In terms of the expansion,
\[
\frac{Q_s(u+i/2)}{C_s} = \prod_k(u +i/2-u_k) = c^{(s)}_s u^s + c^{(s)}_{s-1} u^{s-1} + \ldots c^{(s)}_1 u + c^{(s)}_0,
\]
this ratio is $4 i \, \lambda^2 c^{(s)}_1 / c^{(s)}_0$. 
The series expansion of the hypergeometric function gives
\[
Q_s(u +i/2)={}_2F_1(-s, iu ; 1; 2) = \sum_{k=0}^s \frac{(-s)_k (iu)_k}{(1)_k} \frac{2^k}{k!}, \quad (a)_k=\Gamma(a+k)/\Gamma(a),
\]
which implies
\[
c^{(s)}_0 = 1, \quad c^{(s)}_1   = i \sum_{k=1}^s \frac{(-s)_k (k-1)!  }{(1)_k} \frac{2^k}{ k!} = i \sum_{k=1}^s \frac{s! (-2)^k}{k(s-k)! k!}.
\]
To find $c_1^{(s)}$ we evaluate
\<
c^{(s)}_1 - c^{(s-1)}_1 \eq i \sum_{k=1}^s \frac{s! (-2)^k}{k(s-k)! k!} - i \sum_{k=1}^{s-1} \frac{(s-1)! (-2)^k}{k(s-k-1)! k!} 
\nln
\eq
\frac{i}{s} (-2)^s+ \frac{i}{s}  \sum_{k=1}^{s-1} \frac{s! (-2)^k}{(s-k)! (k)!}
\nln
\eq
\frac{i}{s} ((-1)^s -1),
\>
where we completed the binomial expansion of $(1- 2)^s$ to reach the last line. Since  $c_1^{(0)} =0 $, $c_1^{(s)} = i\Big(S_{-1}(s) - S_1(s) \Big)$ and 
\[
\Delta_{s,2} = 4 i  \frac{c^{(s)}_1}{c^{(s)}_0} = 4  \Big(S_1(s) - S_{-1}(s)\Big).
\]
As $s \rightarrow \infty$, $S_1(s) \rightarrow \log s$ while $S_{-1}(s) \rightarrow - \log 2$, so the cusp anomalous dimension has weak coupling expansion
\[
f(\lambda) =  4 \lambda^2 + \mathcal{O}(\lambda^3) . 
\]
We expect the methods of \cite{Kotikov:2008pv} could be used to obtain the four-loop correction to the twist-one spectrum, up to a currently unknown coefficient from $h(\lambda)$. 
\newpage
\section{Local contribution to $\comm{\gen{Y}^{aa}}{\hat{\gen{Q}}}$ \label{sec:localycomm}} 
In this appendix we complete the proof of $\alg{osp}(4|2)$ Yangian symmetry of Section \ref{sec:yangian} by computing what we called the local term of the commutator between $\hat{\gen{Q}}$ and the $\gen{Y}^{aa}$.
Recall
\<
4 \gen{Y}^{aa} \eq 
\sum_{i < j} 2 \varepsilon_{bc} \gen{R}^{bc}_i \gen{R}^{ca}_j -  \varepsilon_{\mathfrak{bc}} \gen{Q}^{a1 \mathfrak{b}}_i \gen{Q}^{a2 \mathfrak{c}}_j + \varepsilon_{\mathfrak{bc}} \gen{Q}^{a2 \mathfrak{b}}_i \gen{Q}^{a1 \mathfrak{c}}_j
\nln
\eq \gen{Y}^{aa}_{\gen{R}} + \gen{Y}^{aa}_{\gen{Q}^1} + \gen{Y}^{aa}_{\gen{Q}^2}. 
\>
So the local contributions of the form (\ref{eq:local}) from $\gen{Y}^{aa}_{\gen{R}}$ are
\[ 
2 \varepsilon_{bc} \sum_{i} \Big( \gen{R}^{ab}_i (\gen{R}^{ac}_{i+1} +\gen{R}^{ac}_{i+2} ) +\gen{R}^{ab}_{i+1}\gen{R}^{ac}_{i+2} \Big) \gen{\hat{Q}}_i.
\]
Using the expressions for the $\gen{R}$ generators given in (\ref{eq:osp42lightcone}), and the expression for $\hat{\gen{Q}}$,
\< \label{eq:appqhat}
\hat{\gen{Q}} \state{x, \theta, \bar{\eta}} \eq \int_0^x \mathrm{d}y \,  \Big( \varepsilon_{ab} \partial^a_2 \partial^b_3  +  \varepsilon_{\mathfrak{ab}} \partial^{\mathfrak{a}}_2  \partial^{\mathfrak{b}}_3 \Big) \state{x, \theta, \bar{\eta}; y, \bar{\theta}_2, \eta_2;y, \theta_3, \bar{\eta}_3}   
 \nl
 + \int_0^x \mathrm{d}y \,  \Big( \varepsilon_{ab}  \partial^a_1 \partial^b_2   +  \varepsilon_{\mathfrak{ab}}  \partial_1^{\mathfrak{a}} \partial_2^{\mathfrak{b}}  \Big) \state{ y, \theta_1, \bar{\eta}_1; y, \bar{\theta}_2, \eta_2; x, \theta, \bar{\eta}} 
 \nl
  -\varepsilon_{\mathfrak{ab}} \bar{\eta}^{\mathfrak{a}} \partial_2^{\mathfrak{b}} \varepsilon_{cd}  \partial_1^c\partial_3^d    \state{ x, \theta_1, \bar{\eta}_1;x, \bar{\theta}_2, \eta_2; x, \theta_3, \bar{\eta}_3 } ,
   \>
 we obtain
\<
4 \comm{\gen{Y}^{aa}_{\gen{R}}}{\hat{\gen{Q}}}_{\text{local}} \state{x, \theta, \bar{\eta}} \eq  \int_0^x \mathrm{d}\!y \,  2   \partial_2^a  \partial_3^a    \state{x, \theta, \bar{\eta}; y, \bar{\theta}_2, \eta_2;y, \theta_3, \bar{\eta}_3}   
 \nl
 + \int_0^x \mathrm{d}\!y \, 2   \partial_1^a  \partial_2^a   \state{ y, \theta_1, \bar{\eta}_1; y, \bar{\theta}_2, \eta_2; x, \theta, \bar{\eta}} 
 \nl
  - 2  \varepsilon_{\mathfrak{bc}}\bar{\eta}^{\mathfrak{b}} \bar{\partial}_2^{\mathfrak{c}}   \partial_1^a \partial_3^a    \state{ x, \theta_1, \bar{\eta}_1;x, \bar{\theta}_2, \eta_2; x, \theta_3, \bar{\eta}_3 } . \label{eq:rycom}
 \>
This above calculation is simplified using the vanishing of a single $\gen{R}$ applied to  $\gen{R}$-singlet combinations of derivatives.

Next the local contributions from $\gen{Y}^{aa}_{\gen{Q}^1}$ are, after regrouping, 
\[  \label{eq:twoparts}
-  \varepsilon_{\mathfrak{bc}}  \sum_{i} \Big((\gen{Q}^{a1 \mathfrak{b}}_i + \gen{Q}^{a1 \mathfrak{b}}_{i+1})\gen{Q}^{a2 \mathfrak{c}}_{i+2} + \gen{Q}^{a1 \mathfrak{b}}_i \gen{Q}^{a2 \mathfrak{c}}_{i+1}   \Big) \gen{\hat{Q}}_i.
\]
First we compute the action of the first term (including the two supercharge terms in parenthesis) on $\state{x, \theta, \bar{\eta}}$. The three lines of (\ref{eq:appqhat}) yield respectively,
\begin{eqnarray} 
& & \int_0^x \mathrm{d}y \,  \bigg(-\varepsilon_{\mathfrak{bc}} \gen{Q}_1^{a1\mathfrak{b}} \Big(2 y \partial_2^a\bar{\partial}_3^{\mathfrak{c}} +(2 y \partial_{y_3}+1)\bar{\partial}_2^{\mathfrak{c}} \partial_3^a \Big) \label{eq:11}
 \notag \\
 & & -4 \partial_2^a\partial_3^a y \partial_{y_3} -2 \partial_2^a\partial_3^a \bigg)  \state{x, \theta, \bar{\eta}; y, \bar{\theta}_2, \eta_2;y, \theta_3, \bar{\eta}_3} 
 \\
&&-\varepsilon_{\mathfrak{bc}} \partial_1^a \bar{\partial}_2^{\mathfrak{b}}\gen{Q}_3^{a2\mathfrak{c}} \Big(\state{x,\theta_1, \bar{\eta}_1;x, \bar{\theta}_2, \eta_2;x, \theta, \bar{\eta}}-\state{0,\theta_1, \bar{\eta}_1;0, \bar{\theta}_2, \eta_2;x, \theta, \bar{\eta}}\Big) \label{eq:12}
\\
&& +2x \partial_1^a\partial_2^a \varepsilon_{\mathfrak{bc}}\bar{\eta}^{\mathfrak{b}} \bar{\partial}_3^{\mathfrak{c}} \state{ x, \theta_1, \bar{\eta}_1;x, \bar{\theta}_2, \eta_2; x, \theta_3, \bar{\eta}_3 }. \label{eq:13}
 \end{eqnarray}
In the first line  we introduced the notation $\partial_{y_i}$ ($i=3$ in this case), which should be understood as $\partial_y$ acting only on the $i$th site.  Similarly, we compute the action of  the last term of (\ref{eq:twoparts}) on $\state{x, \theta, \bar{\eta}}$, again splitting according to the three lines of $\hat{\gen{Q}}$ in (\ref{eq:appqhat}), yielding
 \begin{eqnarray}
& & \int_0^x \mathrm{d}y \,  \varepsilon_{\mathfrak{bc}} \gen{Q}_1^{a1\mathfrak{b}}\Big(-(2 y\partial_{y_2} +1)\bar{\partial}_2^{\mathfrak{c}}\partial_3^a + 2 y \partial_2^a \bar{\partial}_3^{\mathfrak{c}}\Big)\state{x, \theta, \bar{\eta}; y, \bar{\theta}_2, \eta_2;y, \theta_3, \bar{\eta}_3} \label{eq:21}
\\
& & + \int_0^x \mathrm{d}y \, 4  y \partial_{y_1} \partial_1^a \partial_2^a  \state{ y, \theta_1, \bar{\eta}_1; y, \bar{\theta}_2, \eta_2; x, \theta, \bar{\eta}}  \label{eq:22}
\\
& & + 2  x \varepsilon_{\mathfrak{bc}} \bar{\partial}_1^{\mathfrak{b}} \partial_2^a \partial_3^a \bar{\eta}^{\mathfrak{c}} \state{ x, \theta_1, \bar{\eta}_1;x, \bar{\theta}_2, \eta_2; x, \theta_3, \bar{\eta}_3 }. \label{eq:23}
\end{eqnarray}
It remains to compute the local terms from the last part of $\gen{Y}^{aa}$ (\ref{eq:Ycc}), labeled $\gen{Y}^{aa}_{\gen{Q}^2}$. However, as we noted previously, $\hat{\gen{Q}}$ is odd under spin-chain parity $\mathbf{p}$, and it is straightforward to check that the Yangian generators are as well.  Therefore, we find
\[
\comm{\gen{Y}^{aa}_{\gen{Q}^2}}{\hat{\gen{Q}}}_{\text{local}}= \mathbf{p} \comm{\gen{Y}^{aa}_{\gen{Q}^1}}{\hat{\gen{Q}}}_{\text{local}}\mathbf{p}^{-1},
\]
where $4 \comm{\gen{Y}^{aa}_{\gen{Q}^1}}{\hat{\gen{Q}}}_{\text{local}}$ is given by the sum of (\ref{eq:11}-\ref{eq:13}) and (\ref{eq:21}-\ref{eq:23}). So now we add (\ref{eq:11}-\ref{eq:13}) and (\ref{eq:21}-\ref{eq:23}), their images under $\mathbf{p}$, and (\ref{eq:rycom}) to give the full local contribution to the commutator. Many terms cancel\footnote{
Specifically, the cancellations are: last term of (\ref{eq:11}) with first term of (\ref{eq:rycom}) (and $\mathbf{p}$ image cancellation), first term of (\ref{eq:11}) with second term of (\ref{eq:21}), (\ref{eq:13}) with $\mathbf{p}$  of (\ref{eq:23}) (and $\mathbf{p}$ image), second-to-last term of (\ref{eq:11}) with $\mathbf{p}$ of (\ref{eq:22}) (and $\mathbf{p}$ image).}.  Also, the second term of (\ref{eq:11}) combines with the first term of (\ref{eq:21}) to give 
\< \label{eq:last}
-2 \int_0^x \mathrm{d}y \, \partial_y\Big(y \varepsilon_{\mathfrak{bc}} \gen{Q}_1^{a1\mathfrak{b}} \bar{\partial}_2^{\mathfrak{c}} \partial_3^a \state{x, \theta, \bar{\eta};y, \bar{\theta}_2, \eta_2;y, \theta_3,\bar{\eta}_3} \Big) \eq
\notag \\
 -2 x \varepsilon_{\mathfrak{bc}} \gen{Q}_1^{a1\mathfrak{b}} \bar{\partial}_2^{\mathfrak{c}} \partial_3^a \state{x, \theta, \bar{\eta};x, \bar{\theta}_2, \eta_2;x, \theta_3,\bar{\eta}_3}. & & 
\>
All that remains is the last term of (\ref{eq:rycom}), (\ref{eq:12}) and its $\mathbf{p}$ image, and this last expression (\ref{eq:last}) and its $\mathbf{p}$ image. We write out all these terms, expanding the remaining $\gen{Q}$ factors, yielding
\begin{eqnarray}
& &   - 2  \varepsilon_{\mathfrak{bc}}\bar{\eta}^{\mathfrak{b}} \bar{\partial}_2^{\mathfrak{c}}   \partial_1^a \partial_3^a    \state{ x, \theta_1, \bar{\eta}_1;x, \bar{\theta}_2, \eta_2; x, \theta_3, \bar{\eta}_3 } 
\notag \\ 
& & -\varepsilon_{\mathfrak{bc}} \partial_1^a \bar{\partial}_2^{\mathfrak{b}} (2 x \theta^a \bar{\partial}_3^{\mathfrak{c}} + 2 x \partial_{x_3} \partial_3^a \bar{\eta}^{\mathfrak{c}}  + \partial_3^a \bar{\eta}^{\mathfrak{c}} ) \state{x,\theta_1, \bar{\eta}_1;x, \bar{\theta}_2, \eta_2;x, \theta, \bar{\eta}} - \text{parity}
\notag \\
& & -2 x \varepsilon_{\mathfrak{bc}} (\theta^a \bar{\partial}_1^{\mathfrak{b}} + \partial_{x_1} \partial_1^a \bar{\eta}^\mathfrak{b})  \bar{\partial}_2^{\mathfrak{c}} \partial_3^a \state{x, \theta, \bar{\eta};x, \bar{\theta}_2, \eta_2;x, \theta_3,\bar{\eta}_3} - \text{parity}
\notag \\
& & + \varepsilon_{\mathfrak{bc}} \partial_1^a \bar{\partial}_2^{\mathfrak{b}}\gen{Q}_3^{a2\mathfrak{c}} \state{0,\theta_1, \bar{\eta}_1;0, \bar{\theta}_2, \eta_2;x, \theta, \bar{\eta}} +  \text{parity}
\nln
\eq \varepsilon_{\mathfrak{bc}} \partial_1^a \bar{\partial}_2^{\mathfrak{b}}\gen{Q}_3^{a2\mathfrak{c}} \state{0,\theta_1, \bar{\eta}_1;0, \bar{\theta}_2, \eta_2;x, \theta, \bar{\eta}} - \varepsilon_{\mathfrak{bc}} \gen{Q}_1^{a2\mathfrak{b}} \bar{\partial}_2^{\mathfrak{c}} \partial_3^a \state{x, \theta, \bar{\eta};0, \bar{\theta}_2, \eta_2;0,\theta_3, \bar{\eta}_3}
\notag \\
& \mapsto & \sum_i \varepsilon_{\mathfrak{bc}} \acute{Z}^{a\mathfrak{b}}_i\gen{Q}_{i+1}^{a2\mathfrak{c}} +\sum_i  \varepsilon_{\mathfrak{bc}} \gen{Q}_i^{a2\mathfrak{c}} \grave{Z}^{a\mathfrak{b}}_{i+1}.
\end{eqnarray}
The initial expression simplified as follows. The first line canceled against the terms without $x$ coefficients in the second line, and the remaining terms proportional to $x$ canceled in four pairs, leaving only the fourth line.  The second-to-last line is simply expanding the fourth line (the parity term). We used the symbol $\mapsto$ in the last line for two reasons. We have lifted the one-to-three local site interaction to its homogeneous action on an infinite chain. Also, this expression includes a second set of local actions, on initial sites $\state{x, \bar{\theta}, \eta}.$ For these the calculation is completely analogous.  Starting from the action of $\hat{\gen{Q}}$ on such sites, 
\<
\hat{\gen{Q}} \state{x, \bar{\theta}, \eta} \eq -\int_0^x \mathrm{d}y \,  \Big( \varepsilon_{\mathfrak{ab}} \bar{\partial}^{\mathfrak{a}}_2  \bar{\partial}^{\mathfrak{b}}_3  +  \varepsilon_{ab} \partial^a_2  \partial^b_3   \Big) \state{x, \bar{\theta}, \eta; y, \theta_2, \bar{\eta}_2;y, \bar{\theta}_3, \eta_3}   
 \nl
 - \int_0^x \mathrm{d}y \, \Big( \varepsilon_{\mathfrak{ab}} \bar{\partial}^{\mathfrak{a}}_1  \bar{\partial}^{\mathfrak{b}}_2  +  \varepsilon_{ab} \partial^a_1  \partial^b_2   \Big)  \state{ y, \bar{\theta}_1, \eta_1; y, \theta_2, \bar{\eta}_2; x, \bar{\theta}, \eta} 
 \nl
  +  \varepsilon_{ab} \eta^{a} \partial_2^b \varepsilon_{\mathfrak{cd}}  \bar{\partial}^{\mathfrak{c}}_1 \bar{\partial}^{\mathfrak{d}}_3    \state{ x, \bar{\theta}_1, \eta_1;x, \theta_2, \bar{\eta}_2; x, \bar{\theta}_3, \eta_3 } . 
 \>
 We again can compute the contributions to the local terms of $\gen{Y}_{\gen{R}}$, $\gen{Y}_{\gen{Q}^1}$ and (using $\mathbf{p}$) $\gen{Y}_{\gen{Q}^2}$. After simplification, we find
\<
& &  \varepsilon_{\mathfrak{bc}} \bar{\partial}_1^{\mathfrak{b}} \partial_2^a \gen{Q}^{a 2 \mathfrak{c}}_3 \state{ 0, \bar{\theta}_1, \eta_1; 0, \theta_2, \bar{\eta}_2; x, \bar{\theta}, \eta} + \varepsilon_{\mathfrak{bc}} \gen{Q}^{a 2 \mathfrak{c}}_1  \partial_2^a  \bar{\partial}_3^{\mathfrak{b}} \state{ x, \bar{\theta}, \eta; 0, \theta_2, \bar{\eta}_2; 0, \bar{\theta}_3, \eta_3}
\notag \\
& \mapsto &  \sum_i \varepsilon_{\mathfrak{bc}} \acute{Z}^{a\mathfrak{b}}_i\gen{Q}_{i+1}^{a2\mathfrak{c}} +\sum_i  \varepsilon_{\mathfrak{bc}} \gen{Q}_i^{a2\mathfrak{c}} \grave{Z}^{a\mathfrak{b}}_{i+1}.
\>
As claimed, this gives a precise cancellation with the remaining contribution from bilocal terms of (\ref{eq:bilocalYqhatcomm}).

\section{Projectors in the spin-module representation \label{sec:projectors}}
The results of this section parallel those given for the $\alg{psu}(1,1|2)$ sector of $\mathcal{N}=4$ SYM  spin chain in Appendix E of \cite{Zwiebel:2008gr}. The expression for the dilatation generator (\ref{eq:osp42projectorham}) depends on certain $\alg{osp}(4|2)$ invariant generators. These generators act on pairs of different types of modules, and on pairs of identical modules. First, for the case of different types of modules we introduce the weighted sum of projectors, 
\[
\mathcal{P}(c) = \sum_{i=0}^{\infty} c_i \mathcal{P}^{(i)},
\]
where again $\mathcal{P}^{(i)}$ is the projector for $\alg{osp}(4|2)$ spin  $i$. So for $c_{i'} = \delta_{ii'}$, $\mathcal{P}(c)$ reduces to $\mathcal{P}^{(i)}$. More generally, $\mathcal{P}(c)$ can be written in terms of six component functions that depend on the $c_i$ (we suppress the~argument $c$ of $\mathcal{P}$).
\<
\mathcal{P}\state{\phi_a^{(j)} \psi_b^{(n-j)}} \eq \sum_{k=0}^ n \Big( p_1(n, j, k) \state{\phi_a^{(k)} \psi_b^{(n-k)}} + p_2(n, j, k) \state{\phi_b^{(k)} \psi_a^{(n-k)}} \Big),
\nl
+ \sum_{k=0}^n p_3(n, j, k)  \varepsilon_{ab}\varepsilon^{\mathfrak{cd}} \state{\bar{\psi}_{\mathfrak{c}}^{(k)} \bar{\phi}_{\mathfrak{d}}^{(n-k)}},
\nln
\mathcal{P}\state{\phi_a^{(j)} \bar{\phi}_\mathfrak{b}^{(n-j)}} \eq \sum_{k=0}^ n p_4(n, j, k) \state{\phi_a^{(k)} \bar{\phi}_\mathfrak{b}^{(n-k)}} + \sum_{k=0}^ {n-1} p_5(n, j, k) \state{\bar{\psi}_{\mathfrak{b}}^{(k)} \psi_a^{(n-k-1)}},
\nln
\mathcal{P}\state{\bar{\psi}_{\mathfrak{a}}^{(j)} \psi_b^{(n-j)}} \eq  \sum_{k=0}^n  p_6(n, j, k) \state{\bar{\psi}_{\mathfrak{a}}^{(k)} \psi_b^{(n-k)}} + \sum_{k=0}^ {n+1}  p_5(n+1, k, j) \state{\phi_b^{(k)} \bar{\phi}_\mathfrak{a}^{(n-k+1)}},
\nln
\mathcal{P}\state{\bar{\psi}_{\mathfrak{a}}^{(j)} \bar{\phi}_{\mathfrak{b}}^{(n-j)}} \eq \sum_{k=0}^n \Big(p_1(n, n-j, n-k) \state{\bar{\psi}_{\mathfrak{a}}^{(k)} \bar{\phi}_{\mathfrak{b}}^{(n-k)}} + p_2(n, n-j, n-k) \state{\bar{\psi}_{\mathfrak{b}}^{(k)} \bar{\phi}_{\mathfrak{a}}^{(n-k)}}\Big)
\nl
+ \sum_{k=0}^n p_3(n, k, j) \varepsilon_{\mathfrak{ab}} \varepsilon^{cd} \state{\phi_c^{(k)} \psi_d^{(n-k)}}.
\>
The $p_l$ for $l=1,2$ only differ by a few minus signs,
\begin{gather}
p_l(n, j, k) = \sqrt{\frac{\pi \Gamma(j+\half)\Gamma(k+\half)(2(n-j)+1)!(2(n-k)+1)!}{j!k!}} \sum_{i=0}^n \bigg(  \frac{2^{j+k-2n}(n-i)!}{4(n+i+1)!}  
\notag \\
 {}_3F_2^{\text{reg}}(-\half-i, -i, -j;\half, 1-i-j+n;1)C_l(i) {}_3F_2^{\text{reg}}(-\half-i, -i, -k;\half, 1-i-k+n;1) \bigg), \notag
\end{gather}
\vspace{-1cm}
\<
C_1(i) \eq \phantom{-}\theta(i-1) c_{i-1} + (2 + \delta_{i,0})c_i + c_{i+1}, 
\nln
C_2(i) \eq -\theta(i-1) c_{i-1} + (2 -\delta_{i,0})c_i - c_{i+1}.  \label{eq:definepl}
\> 
The regularized hypergeometric functions are defined as ordinary hypergeometric functions divided by gamma functions, as 
\[
{}_3F_2^{\text{reg}}(a_1, a_2, a_3; b_1, b_2; z) = \frac{{}_3F_2(a_1, a_2, a_3; b_1, b_2; z)}{\Gamma(b_1)\Gamma(b_2)}.
\]
The remaining four component functions can then be written relatively compactly in terms of these two. 
\<
p_4(n, j, k) \eq \frac{\sqrt{2k+1}}{\sqrt{2j+1}} p_1(n, n-j, n-k) + \frac{\sqrt{2k+1}(2n-2k+1)}{\sqrt{2j+1}(2j-2k+1)}p_2(n, n-j, n-k) 
\nl
+ \theta(n-j-1) \frac{2 \sqrt{(j+1)(n-j)(2(n-k)+1)}}{\sqrt{2j+1}(2(j-k)+1)} p_2(n, j+1, k), 
\nln
p_3(n, j, k) \eq -\frac{\sqrt{2j+1}}{\sqrt{2(n-j)+1}} p_1(n, n-j, n-k) + \frac{\sqrt{2k+1}}{\sqrt{2(n-j)+1}} p_4(n, j, k),
\nln
p_5(n, j, k) \eq \theta(n-k-1)\Big( \frac{\sqrt{2(n-j)}}{\sqrt{2k+1}} p_2(n-1, j, k) - \frac{\sqrt{2j}}{\sqrt{2k+1}} p_3(n-1, k, j-1)\Big), 
\nln
p_6(n, j, k) \eq \frac{\sqrt{2k+1}}{\sqrt{2j+1}} p_1(n, j, k) - \frac{\sqrt{2(n-k)+1}}{\sqrt{2j+1}} p_3(n, j, k).
\>
Note that the first (second) term in the expression for $p_5$ should be set to zero when $n=j$ ($j=0$) (as written, they are $0 \times \infty$). For use in (\ref{eq:osp42projectorham}), one only needs to set $c_i = (-1)^i$, or $c_i = S_1(i)$.

For identical representations we instead define $\mathcal{P}(c)$ as
\[
\mathcal{P}(c) = \sum_{i=0}^{\infty} c_{i} \mathcal{P}^{(i-1/2)}.
\]
Recall that in this case the $\alg{osp}(4|2)$ spin is half-integer valued. Now the expansion of $\mathcal{P}(c)$ is
\<
\mathcal{P}\state{\phi_a^{(j)} \phi_b^{(n-j)}} \eq \sum_{k=0}^ n \Big( p_7(n, j, k) \state{\phi_a^{(k)} \phi_b^{(n-k)}} + p_8(n, j, k) \state{\phi_b^{(k)} \phi_a^{(n-k)}} \Big)
\nl
+ \sum_{k=0}^ {n-1}   p_9(n, j, k) \varepsilon_{ab}\varepsilon^{\mathfrak{cd}} \state{\bar{\psi}_{\mathfrak{c}}^{(k)} \bar{\psi}_{\mathfrak{d}}^{(n-k-1)}}
\nln
\mathcal{P}\state{\phi_a^{(j)} \bar{\psi}_\mathfrak{b}^{(n-j)}} \eq \sum_{k=0}^ n \Big(p_{10}(n, j, k) \state{\phi_a^{(k)} \bar{\psi}_\mathfrak{b}^{(n-k)}} +  p_{11}(n, j, k) \state{\bar{\psi}_{\mathfrak{b}}^{(k)} \phi_a^{(n-k)}}\Big),
\nln
\mathcal{P}\state{\bar{\psi}_\mathfrak{a}^{(j)}\phi_b^{(n-j)} } \eq \sum_{k=0}^ n \Big(p_{10}(n, n-j, n-k) \state{\bar{\psi}_\mathfrak{a}^{(k)}\phi_b^{(n-k)} } +  p_{11}(n, n-j, n-k) \state{\phi_b^{k} \bar{\psi}_\mathfrak{a}^{(n-k)}}\Big),
\nln
\mathcal{P}\state{\bar{\psi}_\mathfrak{a}^{(j)}\bar{\psi}_\mathfrak{b}^{(n-j)} } \eq \sum_{k=0}^ n \Big(p_{12}(n, j, k)\state{\bar{\psi}_\mathfrak{a}^{(k)}\bar{\psi}_\mathfrak{b}^{(n-k)} } +  p_{13}(n, j, k)\state{\bar{\psi}_\mathfrak{b}^{(k)}\bar{\psi}_\mathfrak{a}^{(n-k)} } \Big)
\nl
+ \sum_{k=0}^{n+1} p_9(n+1, k, j)  \varepsilon_{\mathfrak{ab}} \varepsilon^{cd}\state{\phi_c^{(k)} \phi_d^{(n-k+1)}}.
\>
$p_{10}$ actually takes the same form as given in (\ref{eq:definepl}), with 
\[
C_{10}(i) = 2 c_i + 2 c_{i+1}.
\]
For $p_m$, $m=7, 8$ there are some shifts of arguments, 
\begin{gather}
p_m(n, j, k) = \sqrt{\frac{\pi \Gamma(j+\half)\Gamma(k+\half)(2(n-j))!(2(n-k))!}{j!k!}} \sum_{i=0}^n \bigg(  \frac{2^{1+j+k-2n}(n-i)!}{4(n+i)!}  
\notag \\
 {}_3F_2^{\text{reg}}(\half-i, -i, -j;\half, 1-i-j+n;1)D_m(i) {}_3F_2^{\text{reg}}(\half-i, -i, -k;\half, 1-i-k+n;1) \bigg), \notag
\end{gather}
\vspace{-1cm}
\<
D_7(i) \eq \phantom{-}\theta(i-1) c_{i-1} + (2- \delta_{i, 0}) c_i + c_{i+1}, 
\nln
D_8(i) \eq -\theta(i-1) c_{i-1} + (2- \delta_{i, 0})c_i -  c_{i+1}.
\> 
The remaining four component functions again can be written in terms of ones defined earlier.
\<
p_9(n, j, k) \eq \theta(n-k-1)\Big( \frac{\sqrt{2(k+1)}}{\sqrt{2(n-k-1)+1}} p_7(n, k+1, j) 
\nl
- \frac{\sqrt{2j}}{\sqrt{2(n-k-1)+1}} p_{10}(n-1, n-k-1, n-j) \Big), 
\nln
p_{11}(n, j, k) \eq \frac{\sqrt{(n-j+1)}}{\sqrt{k+1}} p_8(n+1, n-j+1, n-k) 
\nl
+ \frac{\sqrt{2j+1}}{\sqrt{2(k+1)}} p_9(n+1, n-k, n-j), 
\nln
p_{12}(n, j, k) \eq \frac{\sqrt{2(n-k)+1}}{\sqrt{2(n-j)+1}}p_{10}(n, n-j, n-k) + \frac{\sqrt{2(j+1)}}{\sqrt{2(n-j)+1}} p_9(n+1, j+1, k)
\nln
p_{13}(n, j, k) \eq -\frac{\sqrt{2k+1}}{\sqrt{2(n-j)+1}} p_{11}(n,n-j, n-k) - \frac{\sqrt{2(j+1)}}{\sqrt{2(n-j)+1}} p_9(n+1, j+1, k).
\nl
\>
Note that the second term in the expression for $p_9$ should be set to zero for $j=0$.
For the conjugate case of $\mathcal{V}_{\bar \phi}^{(4|2)}$, simply remove (add) a bar from (to) all (un)barred module elements, keeping the same component functions. For  (\ref{eq:osp42projectorham}), in these cases we need to evaluate $\mathcal{P}(c)$ only for $c_i = S_1(i-1/2)$.

\section{Unique lift from $\alg{sl}(2)$ to  $\alg{osp}(4|2)$ \label{sec:maptosl2}}
The general argument of Section \ref{sec:uniquelift} can be straightforwardly applied to lift the Hamiltonian uniquely from the $\alg{sl}(2)$ sector to the $\alg{osp}(4|2)$ sector. However,  we still need to  construct invertible maps between the $\alg{osp}(4|2)$ sector and the $\alg{sl}(2)$ sector, as done between the $\alg{osp}(6|4)$ and $\alg{osp}(4|2)$ sectors in Section (\ref{sec:invertiblemap}). We complete this construction here.  Again, simply constructing the map from the larger to smaller sector will be sufficient. Invertibility will follow, as in the last paragraph of Section \ref{sec:invertiblemap}. 
 
The $\alg{osp}(4|2)$ raising generators are $\gen{R}^{22}$, $\tilde{\gen{R}}^{22}$, $\gen{J}^{22}$,  and $\gen{Q}^{a2\mathfrak{b}}$. Let $\state{\Omega_I}$ be a linearly-independent basis of three-site highest-weight $\alg{osp}(4|2)$ sector states with identical Cartan charges. Of course, we need not consider $\state{\Omega_I}$ already in the $\alg{sl}(2)$ sector (actually there are no such three-site $\alg{osp}(4|2)$ highest-weight states). So, first assume the $\state{\Omega_I}$ transform with respect to $\alg{su}(2)_{\gen{R}} \otimes \alg{su}(2)_{\tilde{\gen{R}}}$ as
$(\mathbf{3} ,\mathbf{2}),$ or equivalently that the $\state{\Omega_I}$ have  $\alg{su}(2)$ Cartan eigenvalues $[R^{12}, \tilde{R}^{12}]$ = $[-1, -1/2]$ (The third Cartan charge is unimportant for this section).  Then consider 
\[
\gen{Q}^{211}\state{\Omega_I}.
\]
These states have ($\alg{su}(2)$) Cartan charges $[-3/2, 0]$ and, equivalently, are in the $\alg{sl}(2)$ subsector. However, a priori the $\gen{Q}^{211}\state{\Omega_I}$ could be zero. But if they were zero, then the highest-weight states $\state{\Omega_I}$ would satisfy a second $1/12$ BPS condition with respect to $\gen{Q}^{211}\sim \gen{Q}_{14,1}$ and $\gen{Q}^{122}\sim \gen{S}^{14,1}$.  This BPS condition(s) is inconsistent (for any possible value of $J^{12}$) with  the $\alg{su}(2)$ Cartan charges of the $\state{\Omega_I}$.  Therefore, $\gen{Q}^{211}\state{\Omega_I}$ is nonvanishing and in this case $\gen{Q}^{211}$ gives the invertible map to the $\alg{sl}(2)$ sector.   

Up to interchanging $\alg{su}(2)_{\gen{R}}$ and $\alg{su}(2)_{\tilde{\gen{R}}}$ charges, the only other possibility for the Cartan charges of the $\state{\Omega_I}$ is
$[R^{12}, \tilde{R}^{12}]$ = $[-1/2, 0]$. Now descendants in the $\alg{sl}(2)$ sector are given by
\[
\gen{Q}^{212}\gen{Q}^{211} \state{\Omega_I} = -\gen{Q}^{211} \gen{Q}^{212}\state{\Omega_I},
\]
where the equality follows from the vanishing anticommutator between these supercharges. The same BPS condition as in the previous case would apply if the $\gen{Q}^{211} \state{\Omega_I}$ were zero, and  this is inconsistent with Cartan charges $[-1/2, 0]$.  Therefore, we only need to show that it is impossible to simultaneously satisfy
\[
\gen{Q}^{212}\gen{Q}^{211} \state{\Omega_I} =0 \quad \text{and} \quad \gen{Q}^{211} \state{\Omega_I} \neq 0.
\]
$\gen{Q}^{211} \state{\Omega_I} $ have Cartan charges $[-1, 1/2]$. Also, they are annihilated by  $\gen{Q}^{121} =(\gen{Q}^{212})^\dagger $ because of the anticommutation relation (\ref{eq:osp42anticom}) and because $\tilde{\gen{R}}^{11}$ annihilates the $\state{\Omega_I}$ (they carries no $\tilde{\gen{R}}$ charge). Therefore, if $\gen{Q}^{212}$ annihilated the $\gen{Q}^{211} \state{\Omega_I} $,  $\gen{Q}^{211} \state{\Omega_I} $ would satisfy another BPS condition, which again turns out to be incompatible with their Cartan charges. We conclude that in this case $\gen{Q}^{212}\gen{Q}^{211} \state{\Omega_I}$ gives nonvanishing descendants in the $\alg{sl}(2)$ sector, and therefore we can always construct the required invertible maps between the the $\alg{osp}(4|2)$ sector and a $\alg{sl}(2)$ sector.


\end{document}